\documentclass[fleqn,11pt]{wlscirep}
\usepackage[utf8]{inputenc}
\usepackage[T1]{fontenc}
\usepackage{hyperref}
\usepackage{authblk}
\usepackage{xcolor}
\usepackage{ragged2e}
\usepackage{algorithm}
\usepackage{algpseudocode}
\usepackage{multirow}
\usepackage{cleveref}

\titlecontents{subsection}[4pc]
  {\addvspace{2pt}\small\sffamily}
  {\contentslabel[\thecontentslabel]{2.5pc}}
  {}
  {\ \titlerule*[.5pc]{.}\ \thecontentspage}
  []

\definecolor{R0}{rgb}{0,0,255}

\title{OMNIA: Universal Neural Samplers for Atomistic Simulation}
\author[1,2,3,*,$\#$]{Mouyang Cheng}
\author[1,4,*,$\#$]{Denis Blessing}
\author[1,5,$\#$]{Jiajun He}
\author[2,6]{Mingda Li}
\author[1,$\dagger$]{Carles Domingo-Enrich}
\author[1,$\dagger$]{Yuanqi Du}

\affil[1]{Microsoft Research New England, Cambridge, MA 02142, USA}
\affil[2]{Center for Computational Science and Engineering, MIT, Cambridge, MA 02139, USA}
\affil[3]{Department of Materials Science and Engineering, MIT, Cambridge, MA 02139, USA}
\affil[4]{Karlsruhe Institute of Technology, 76131 Karlsruhe, Germany}
\affil[5]{University of Cambridge, Cambridge, CB2 1PZ, UK}
\affil[6]{Department of Nuclear Science and Engineering, MIT, Cambridge, MA 02139, USA}

\affil[*]{These authors contributed equally.}
\affil[$\#$]{Work done during internship at Microsoft Research.}
\affil[$\dagger$]{Correspondence. Email: carlesd@microsoft.com, yuanqidu@microsoft.com}

\begin{abstract}
A central goal of atomistic simulation is to predict the equilibrium behavior of materials by sampling thermodynamic ensembles. Yet large free energy barriers can prevent conventional simulations from exploring relevant configurations and determining their equilibrium probabilities. 
Here we introduce OMNIA, a universal neural framework for equilibrium sampling across thermodynamic ensembles. By combining conditional sampling with thermodynamic weights, OMNIA unifies ensembles with fixed or fluctuating volume, composition and particle number, encompassing $NVT$, $NPT$, $\Delta\mu NPT$ and $\mu VT$. 
OMNIA learns diffusion processes that transport samples from simple prior distributions to the target ensemble. The associated path weights enable the estimation of thermodynamic properties, including free energy, as well as conditional steering at inference time.
Applied to oxide glasses, polymorphs of SiO$_2$ and Ti, Cu-Ag and Cu-Ni alloys, hydrogen adsorption and absorption on Pt(111) and Pd(111), and sodium storage in hard carbon, OMNIA reproduces structural distributions, free energy landscapes, phase equilibria and particle uptake, while amortizing sampling across compositions, thermodynamic conditions and host structures. 
These capabilities establish a foundation for scalable atomistic simulation of realistic materials, opening new possibilities for inverse design and materials discovery.
\end{abstract}

\begin{document}

\addtocontents{toc}{\protect\setcounter{tocdepth}{-10}}

\flushbottom
\maketitle
\thispagestyle{empty}

\section*{Introduction}
Since early molecular dynamics (MD) studies of hard-sphere systems and simple liquids, atomistic simulation has become a foundation for understanding the structure, dynamics and properties of matter\cite{alder1957phase,rahman1964correlations,tuckerman2023statistical}. 
Over subsequent decades, advances in algorithms and high-performance computing have extended atomistic simulations to hundreds of millions of atoms\cite{jia2020pushing,casalino2021ai}, while machine learning interatomic potentials (MLIPs) have approached first-principles accuracy across increasingly broad chemical space\cite{unke2021machine,batatia2025foundation,wood2026family}. 
Yet even an exact interatomic potential would not solve the equilibrium problem, which requires determining the relative probabilities of relevant configurations. Predicting stable phases, chemical ordering in alloys or the response of surfaces to their environment therefore remains a sampling problem.
MD and conventional Monte Carlo (MC) estimate these distributions through successive local transitions in configuration space\cite{metropolis1953equation,frenkel2023understanding,tuckerman2023statistical}. When relevant states are separated by large free energy barriers, equilibration becomes fundamentally difficult because the required rare transitions can occur far beyond accessible simulation time scales. 
Enhanced sampling methods such as umbrella sampling and metadynamics accelerate these transitions through biasing along selected collective variables\cite{torrie1977nonphysical,laio2002escaping}, but their efficiency can deteriorate when suitable collective variables are difficult to identify.

Neural samplers provide an alternative by learning transport from a simple prior to a target thermodynamic distribution. Boltzmann generators pioneered this approach with normalizing flows, combining the generation of independent configurations with likelihood-based reweighting to correct sampling bias~\cite{noe2019boltzmann}. More recently, diffusion-based neural samplers have scaled this approach to realistic atomistic systems while learning directly from energy functions, without requiring equilibrium training configurations~\cite{cheng2026atlas,blessing2026janus}. 
Conditioning further allows models to share the cost of learning across thermodynamic states~\cite{lee2026predicting}. Beyond generating configurations, neural samplers can further provide free energy estimates and enable conditional steering at inference time.
Despite this progress, a broader challenge is to connect neural samplers across the thermodynamic constraints encountered in materials. Specialized methods have addressed isothermal-isobaric systems, semi-grand-canonical alloys and grand-canonical sampling of on-lattice defects separately~\cite{van2023boltzmann,schebek2024efficient,damewood2022sampling,cheng2026atlas,blessing2026janus}. 
Yet these ensembles share a common statistical structure: conditional distributions at fixed thermodynamic variables are combined according to the thermodynamic weights of those variables. For example, an $NPT$ ensemble combines configurations sampled at fixed volume with the thermodynamic weight of each volume, with analogous constructions for composition in $\Delta\mu NPT$ and particle number in $\mu VT$.

Here we introduce OMNIA, a universal neural sampler that exploits this structure to unify $NVT$, $NPT$, $\Delta\mu NPT$ and $\mu VT$ sampling, spanning the principal thermodynamic degrees of freedom in equilibrium atomistic simulation: structure, volume, composition and particle number~\cite{frenkel2023understanding,tuckerman2023statistical}. Diffusion models learn conditional distributions directly from interatomic potentials, while thermodynamic marginals determine how these distributions are combined across thermodynamic conditions.
Free energy estimates determine the relative probabilities of competing phases, allowing them to be sampled independently without crossing the barriers between them through local moves.
We demonstrate OMNIA across ordered and disordered materials, competing phases and heterogeneous interfaces. 
Composition-conditioned sampling generates multicomponent oxide glasses, while phase-resolved sampling determines polymorphic phase equilibria in SiO$_2$ and Ti and the phase diagrams in Cu-Ag and Cu-Ni alloys. Grand-canonical sampling reproduces hydrogen adsorption and absorption on Pt(111) and Pd(111), as well as sodium storage in hard carbon anodes. 
Across these applications, learned OMNIA samplers are reused over diverse compositions, thermodynamic conditions and host material structures, substantially reducing the need for repeated system-specific simulations. 
OMNIA thus provides a unified framework for scalable atomistic simulation across diverse materials, phases and thermodynamic ensembles.

\section*{Results}
\subsection*{A unified framework for thermodynamic sampling}
Starting from a chosen reference ensemble, one can construct additional thermodynamic ensembles by allowing previously fixed system variables to fluctuate under imposed reservoir conditions. 
Let $z$ denote the variables released in this construction, $x$ the degrees of freedom already allowed to fluctuate in the conditional ensemble, and $\lambda$ the fixed thermodynamic conditions defining the target ensemble. 
At fixed $z$, the conditional distribution $p_\lambda(x\mid z)$ corresponds to the conditional ensemble and the target distribution $p_\lambda(x,z)$ can be written as
\begin{equation}
p_\lambda(x,z)
=p_\lambda(x\mid z)\,p_\lambda(z),
\qquad
p_\lambda(z)\propto {Z_\lambda}(z){e^{-\beta R_\lambda(z)}},
\label{eq:ensemble_composition}
\end{equation}
where $\beta = (k_BT)^{-1}$, $Z_\lambda(z)$ is the partition function of the reference ensemble at fixed $z$, including its existing thermodynamic couplings, and $R_\lambda(z)$ is the additional reservoir contribution associated with releasing $z$.

\begin{figure}[!t]
  \centering
  \includegraphics[width=0.95\textwidth]{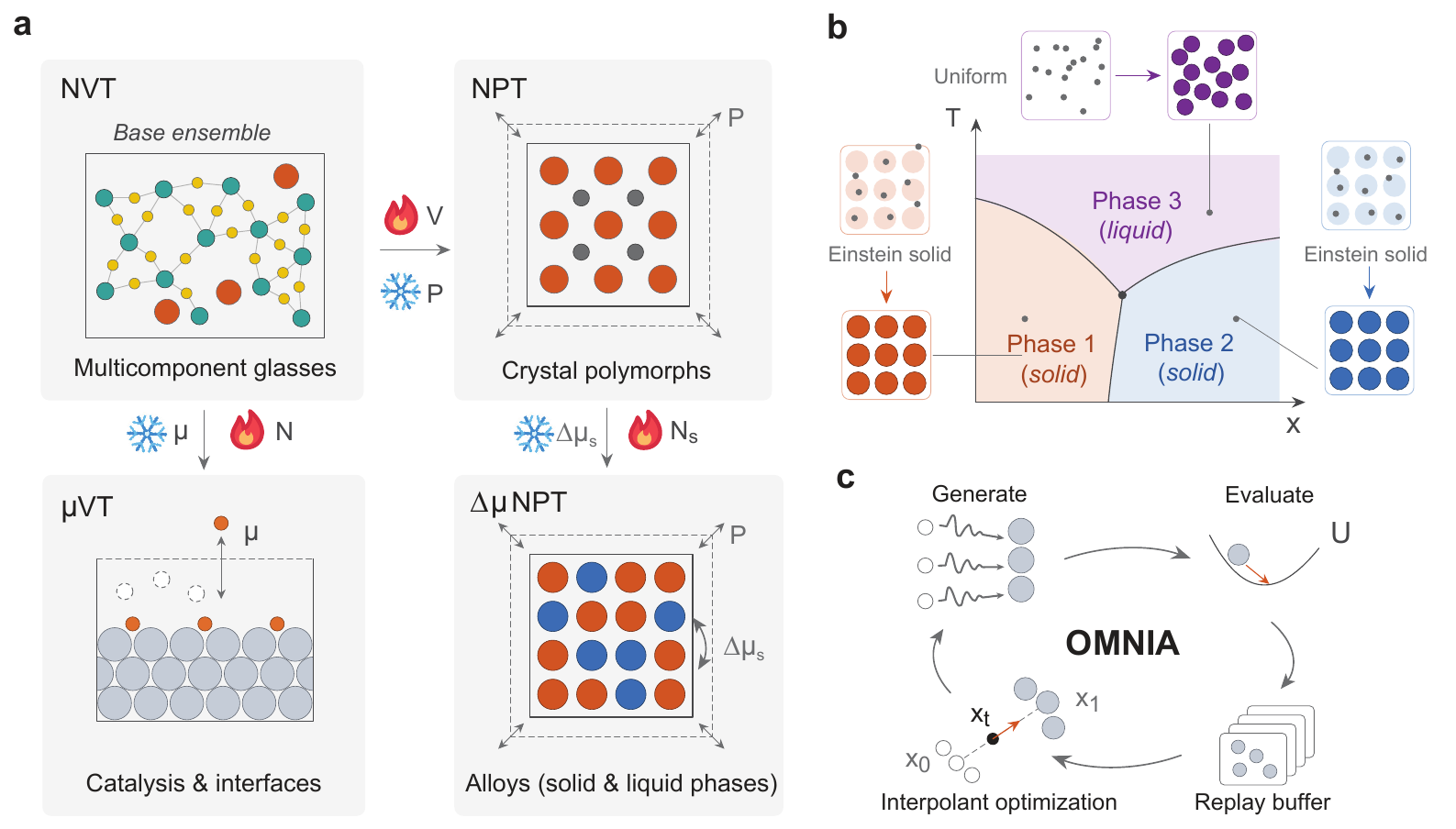}
  \caption{\textbf{Overview of OMNIA.}
\textbf{a.} Unified sampling across thermodynamic ensembles. Starting from the canonical $NVT$ ensemble, OMNIA allows volume $V$, particle number $N$ or species populations $N_s$ to fluctuate under their conjugate thermodynamic controls $P$, $\mu$ or $\Delta\mu_s$, respectively. Flame and snowflake symbols indicate released variables and externally imposed conditions.
\textbf{b.} Sampling across competing thermodynamic phases. Separate diffusion samplers generate configurations within each phase from phase-specific priors, illustrated here by uniform priors for liquids and Einstein solids for crystalline phases. Free energy estimates determine their phase boundaries.
\textbf{c.} Self-bootstrapped training of OMNIA. Configurations are iteratively generated, evaluated using an interatomic potential and stored in a replay buffer. At each iteration, the sampler is refined by learning along interpolants $x_t$ between prior samples $x_0$ and generated configurations $x_1$, without requiring precomputed equilibrium data.}
\label{fig1}
\end{figure}

Taking the canonical $NVT$ ensemble as the conditional ensemble (Fig.\,\ref{fig1}a) gives $R_\lambda(z)=PV$ for $NPT$, $R_\lambda(z)=PV-\sum_{s\neq r}\Delta\mu_sN_s$ for $\Delta\mu NPT$, and $R_\lambda(z)=-\sum_s\mu_sN_s$ for $\mu VT$. Here, $\Delta\mu_s=\mu_s-\mu_r$ is defined relative to a reference species $r$, and the total particle number remains fixed in $\Delta\mu NPT$.
The same construction can be applied successively through intermediate ensembles. For example, taking $NPT$ as the reference and releasing composition gives $\Delta\mu NPT$ with $R_\lambda(z)=-\sum_{s\neq r}\Delta\mu_sN_s$. In this case, $x$ includes the fluctuating volume, and the pressure contribution is already incorporated into $Z_\lambda(z)$. OMNIA implements this framework by either directly learning the target ensemble $p_\lambda(x,z)$ or instead learning the conditional ensemble $p_\lambda(x\mid z)$ and estimating the thermodynamic weights $p_\lambda(z)$.

Competition between phases enters through the same composition principle, as shown in Fig.\,\ref{fig1}b. Partitioning configuration space into regions associated with phases $\alpha$ gives
\begin{equation}
p_\lambda(x,z)
=\sum_\alpha p_{\lambda,\alpha}(x,z)\,p_\lambda(\alpha),
\qquad
p_\lambda(\alpha)\propto e^{-\beta\Phi_\alpha(\lambda)},
\label{eq:phase_ensemble_composition}
\end{equation}
where $p_{\lambda,\alpha}$ is the ensemble restricted to phase $\alpha$ and $\Phi_\alpha$ is its thermodynamic potential.
OMNIA represents each phase with a separate conditional sampler, or ``expert'', initialized within the corresponding basin, separating sampling within a phase from the much slower transitions between phases.

OMNIA learns these thermodynamic distributions using diffusion processes that transport simple priors to atomistic configurations~\cite{cheng2026atlas,blessing2026janus}, as shown in Fig.\,\ref{fig1}c. 
Continuous dynamics generate atomic positions and cell degrees of freedom, while masked discrete diffusion generates chemical identities. Training proceeds iteratively from configurations generated by the current model: the interatomic potential provides energy, forces and virials to guide the transport learning. Because the Boltzmann distribution is a fixed point of this procedure, no precomputed equilibrium dataset is required. Path importance weights from the learned stochastic transport provide unbiased estimates of observables and free energy, and enable flexible steering at inference time (Methods).

The unified construction of OMNIA therefore brings problems usually treated separately, including amorphous structure and disorder, structural transformations, alloy phase diagrams, solid-liquid coexistence and interfacial adsorption, within a common atomistic sampling framework. 
Beyond equilibrium sampling, OMNIA can be steered toward target structures, compositions and properties, enabling free energy barrier estimation, inverse design and materials discovery under increasingly complex conditions. 
Below, we demonstrate OMNIA across the atomic structures of oxide glasses in $NVT$, competing polymorphs of SiO$_2$ and Ti in $NPT$, alloy phase diagrams of Cu-Ni and Cu-Ag in $\Delta\mu NPT$, hydrogen adsorption and absorption on catalytic surfaces and sodium storage in hard-carbon anodes in $\mu VT$.

\subsection*{Sampling amorphous structures and crystalline polymorphs}
\begin{figure}[!tp]
  \centering
  \includegraphics[width=\textwidth]{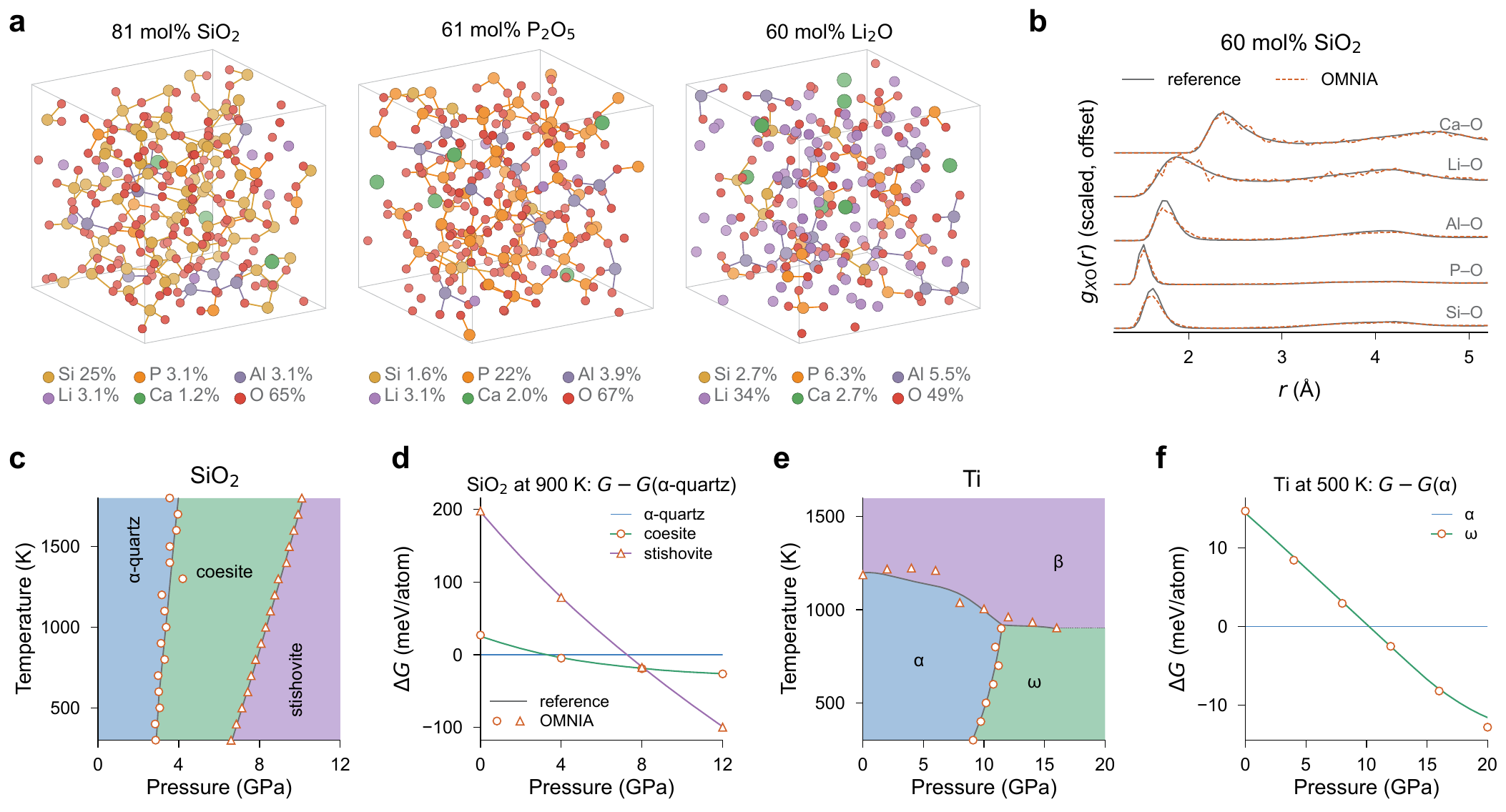}
  \caption{\textbf{OMNIA sampling of multicomponent oxide glasses ($NVT$ ensemble) and competing crystalline polymorphs ($NPT$ ensemble).} 
  \textbf{a.} Representative oxide glass structures generated by a single OMNIA sampler in the $NVT$ ensemble at 1200 K, spanning the SiO$_2$-P$_2$O$_5$-Al$_2$O$_3$-Li$_2$O-CaO composition space. Three compositions enriched in SiO$_2$, P$_2$O$_5$ and Li$_2$O are shown, with corresponding elemental fractions indicated. 
  \textbf{b.} Partial radial distribution functions $g_{\text{XO}}(r)$ ($\text{X}=\text{Ca, Li, Al, P, Si}$) for 60 mol\% SiO$_2$ glass, comparing OMNIA-generated configurations with reference melt-quench simulations. 
  \textbf{c.} Pressure-temperature phase diagram of SiO$_2$ predicted by OMNIA in the $NPT$ ensemble, capturing the competition between $\alpha$-quartz, coesite and stishovite. 
  \textbf{d.} Gibbs free energy differences of the SiO$_2$ polymorphs relative to $\alpha$-quartz at 900 K as a function of pressure. 
  \textbf{e.} Pressure-temperature phase diagram of Ti, showing the competing $\alpha$, $\omega$ and $\beta$ phases. 
  \textbf{f.} Gibbs free-energy difference between the $\omega$ and $\alpha$ phases of Ti at 500 K as a function of pressure. 
  For \textbf{c-f}, OMNIA amortizes sampling across pressures and temperatures for each polymorph, and OMNIA predictions (markers) are compared with reference parallel tempering Monte Carlo (PT-MCMC) results (curves and shading).}
  \label{fig2}
\end{figure}

Oxide glasses and crystalline polymorphs represent two complementary applications of atomistic simulation: resolving the structural diversity of disordered networks across compositions and determining the stability of competing crystalline phases under varying pressure and temperature.
We use OMNIA to investigate these two settings in the $NVT$ and $NPT$ ensembles, respectively, as illustrated in Fig.~\ref{fig2}.

The properties of oxide glasses are strongly controlled by composition, underpinning applications ranging from optical materials~\cite{shearer2023critical} and ion-conducting glasses for energy storage~\cite{wang2023glass} to bioactive materials~\cite{piatti2024tailoring}. 
Yet their atomistic structures are conventionally generated by melt-quench molecular dynamics, requiring an independent cooling trajectory for every composition. Such calculations are expensive, provide no reusable representation of the underlying structural distribution, and offer essentially no amortization across composition space.
OMNIA instead learns a single composition-conditioned distribution across a broad multicomponent chemical space, turning the repeated construction of individual glass structures into a reusable generative model that shares information across compositions and samples new structures without repeating the melt-quench procedure.
We study multicomponent glasses in the SiO$_2$-P$_2$O$_5$-Al$_2$O$_3$-Li$_2$O-CaO composition space using the Bertani-Menziani-Pedone interatomic potentials~\cite{bertani2021improved}.
Trained across this composition space and evaluated on unseen compositions, OMNIA generates continuous oxide networks spanning silica-rich to phosphate- and lithium-rich glasses (Fig.~\ref{fig2}a), with cation-oxygen partial radial distribution functions $g_\text{XO}(r)$ closely matching reference melt-quench simulations (Fig.~\ref{fig2}b).
Moreover, training this reusable sampler requires 24-fold fewer energy evaluations than a single reference melt-quench trajectory at one composition.

Predicting polymorphic phase transitions presents a different challenge, as small differences in Gibbs free energy govern the stability of competing crystalline structures across pressure and temperature.
We use OMNIA to investigate this competition in SiO$_2$ and Ti, two canonical systems with extensively characterized pressure-temperature phase behavior~\cite{akaogi1995thermodynamic,ono2017precise,zhang2008experimental,dewaele2015high}.
We consider $\alpha$-quartz, coesite and stishovite in SiO$_2$, described by the universal MLIP MACE-MH-1~\cite{batatia2025cross}, and the $\alpha$ (hcp), $\omega$ and $\beta$ (bcc) phases of Ti, described by MACE-MPA-0~\cite{batatia2025foundation}. Moving from $NVT$ to $NPT$ releases the fixed-volume constraint and introduces its conjugate field, the pressure $P$.
Following Eq.~\ref{eq:phase_ensemble_composition}, OMNIA assigns one amortized expert to each crystalline phase, conditioned on $T$ and $P$, and determines phase stability from crossings of their Gibbs free energies estimated along the learned diffusion paths (Methods).
As shown in Fig.~\ref{fig2}c-f, OMNIA closely reproduces the phase behavior obtained from MC reference calculations for both systems (Methods).
For SiO$_2$, the predicted quartz-coesite and coesite-stishovite phase boundaries agree with the reference to within a small fraction of a gigapascal, while the Gibbs free energy differences have sub-meV/atom errors on average.
For Ti, OMNIA similarly captures the $\alpha$-$\omega$ boundary and the $\alpha$-$\omega$-$\beta$ triple point, with free energy errors of approximately 1\,meV/atom.
Further quantitative analyses of the accuracy of OMNIA, including the more sensitive $\alpha$-$\beta$ boundary and the improvement from sequential Monte Carlo (SMC), are given in Supplementary Section~\ref{si:npt}.

\subsection*{Phase diagrams of alloys across solid and liquid phases}
Alloy phase diagrams arise from the competition between chemical mixing, structural stability and melting, making their prediction a central challenge in materials thermodynamics.
Resolving these equilibria requires comparing the free energies of competing solid and liquid phases across temperature and composition, which underpins alloy design and processing~\cite{kattner2016calphad}.
The semi-grand-canonical $\Delta\mu NPT$ ensemble provides a natural description, allowing composition to fluctuate in response to the chemical potential difference $\Delta\mu$ while maintaining a fixed total number of atoms.
Conventional MC requires repeated simulations across temperatures and chemical potentials for each competing phase~\cite{sadigh2012scalable,frenkel2023understanding}, while previous neural samplers of chemical disorder have been restricted to crystal lattices~\cite{blessing2026janus}, excluding melting and solid-liquid coexistence.

We use OMNIA to investigate two binary alloys, including their liquid phases at high temperature, with distinct phase behavior: Cu-Ni, which forms a continuous fcc solid solution, and Cu-Ag, which exhibits solid-phase immiscibility and eutectic solidification~\cite{fischer2019systematic,williams2006embedded}.
Moving from $NPT$ to $\Delta\mu NPT$ releases the chemical composition as an additional degree of freedom, which OMNIA samples through discrete diffusion of chemical identities, jointly with continuous diffusion of atomic positions and volume (Methods).
Following Eq.~\ref{eq:phase_ensemble_composition}, expert models on separate phases are conditioned on $T$ and $\Delta\mu$, representing the fcc solid and liquid in Cu-Ni, and the two compositionally distinct fcc solids and liquid in Cu-Ag.
The solid and liquid experts in OMNIA are initialized from an Einstein-crystal reference and a uniform positional prior, respectively, enabling the sampling of both crystalline and liquid structures within the same framework.
Phase boundaries between different phases are then determined by crossings of their semi-grand potentials, reconstructed from the sampled compositions across chemical potentials (Methods).

\begin{figure}[!t]
  \centering
  \includegraphics[width=\textwidth]{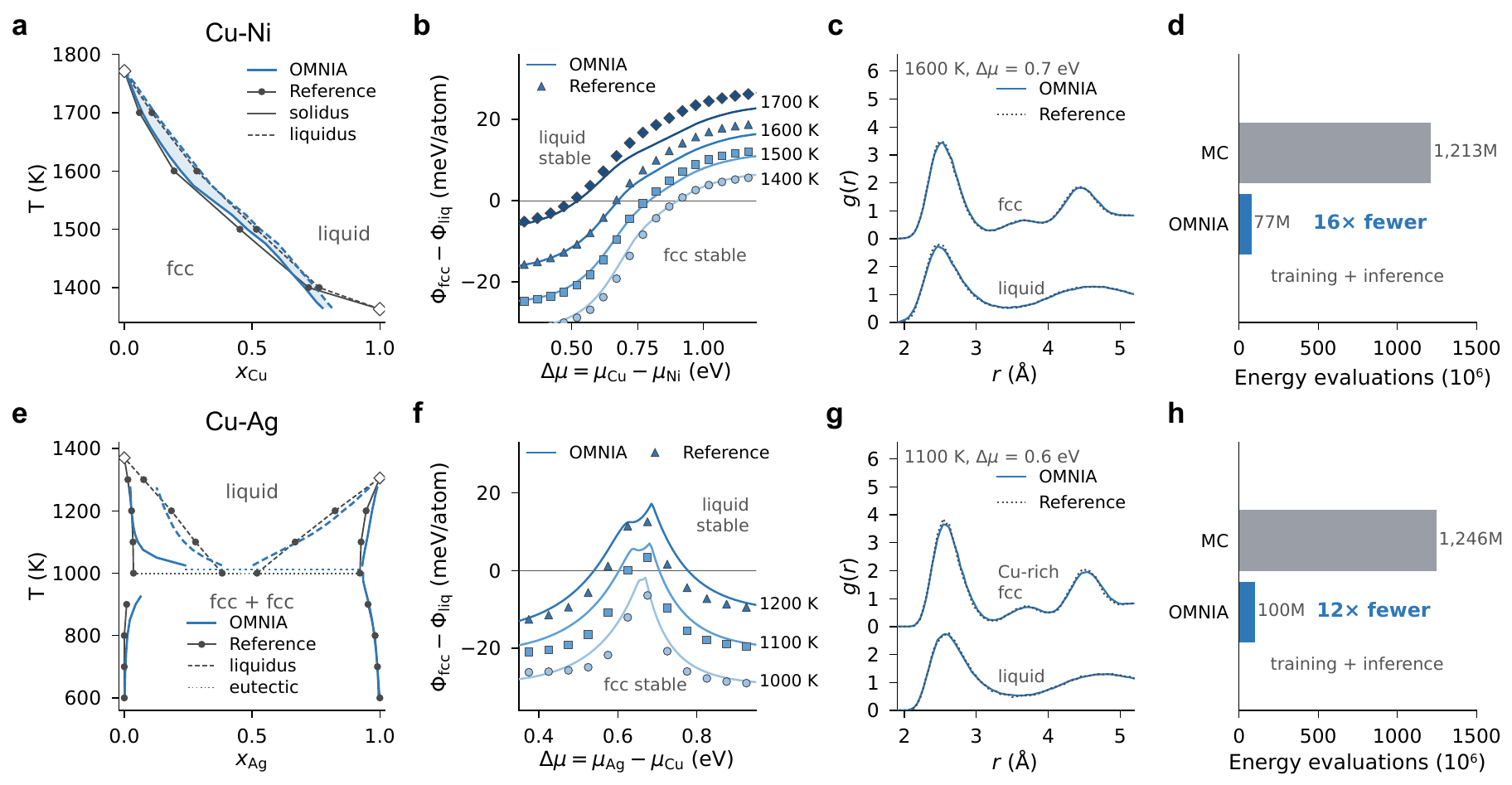}
    \caption{\textbf{OMNIA prediction of phase diagrams in binary alloys in the semi-grand-canonical ensemble.}
    \textbf{a, e.} Temperature-composition phase diagrams of Cu-Ni (\textbf{a}) and Cu-Ag (\textbf{e}), comparing OMNIA predictions with reference Monte Carlo (MC) results, including solid-liquid coexistence and, for Cu-Ag, solid-phase immiscibility and eutectic behavior.
    \textbf{b, f.} Semi-grand-potential differences per atom between fcc solid and liquid phases, $\Phi_{\mathrm{fcc}}-\Phi_{\mathrm{liq}}$, as functions of the chemical-potential difference $\Delta\mu$ at selected temperatures for Cu-Ni and Cu-Ag. Zero crossings determine the solid-liquid phase boundaries.
    \textbf{c, g.} Radial distribution functions $g(r)$ of the solid and liquid phases for Cu-Ni and Cu-Ag, comparing OMNIA with MC references at the indicated thermodynamic conditions.
    \textbf{d, h.} Computational costs measured by the total number of energy evaluations, including both training and inference for OMNIA, compared to MC for Cu-Ni and Cu-Ag, respectively.}
  \label{fig3}
\end{figure}

The resulting alloy phase diagrams are shown in Fig.~\ref{fig3}.
For Cu-Ni, OMNIA reproduces the narrow solidus-liquidus melting lens obtained from reference MC calculations (Fig.~\ref{fig3}a), with compositions on the coexistence line agreeing within 3\%.
The corresponding semi-grand-potential differences closely follow the reference across temperatures and chemical potentials, with mean deviations of only 1-3\,meV/atom (Fig.~\ref{fig3}b).
Beyond phase stability, OMNIA also captures the distinct atomic correlations of the fcc solid and liquid, reproducing their radial distribution functions (Fig.~\ref{fig3}c).
Cu-Ag presents a more complex phase landscape, where two compositionally distinct fcc solids compete with the liquid, giving rise to a broad miscibility gap and eutectic behavior.
OMNIA captures the solid-solid coexistence, both liquidus branches and the eutectic near 1000\,K, in overall agreement with reference MC calculations (Fig.~\ref{fig3}e).
The predicted semi-grand-potential differences reproduce the relative stability of the competing phases across temperatures (Fig.~\ref{fig3}f), while the generated solid and liquid structures closely match reference radial distribution functions (Fig.~\ref{fig3}g).
Further quantitative comparisons, including deviations near the sensitive eutectic and liquidus boundaries, are provided in Supplementary Section~\ref{si:alloys}.

Importantly, the amortization of OMNIA across temperature and chemical potential substantially reduces the computational cost of constructing these phase diagrams.
Including both training and inference, OMNIA requires 16-fold and 12-fold fewer energy evaluations than reference MC for Cu-Ni and Cu-Ag, respectively (Fig.~\ref{fig3}d,h).
Once trained, each expert can be sampled across its thermodynamic window at negligible additional cost, enabling finer resolution of the phase boundaries without repeating independent simulations.

\subsection*{Adsorption and absorption on catalytic surfaces}
Particle exchange with the environment plays a central role in materials applications ranging from catalysis to energy storage and separation. 
We therefore use OMNIA to investigate such open atomistic systems, focusing on hydrogen uptake in transition metals.
Hydrogen adsorption on transition-metal surfaces is central to electrocatalysis and hydrogenation, with platinum among the prototypical catalysts for the hydrogen evolution reaction~\cite{norskov2005trends}. 
Hydrogen on Pt(111) has consequently been studied extensively by both surface experiments and first-principles calculations~\cite{christmann1976adsorption,yan2018hydrogen}. 
Palladium provides a complementary and more complex setting: beyond surface adsorption, hydrogen readily enters the subsurface and bulk to form Pd-H phases, a process relevant to hydrogen storage, separation and catalytic hydrogenation~\cite{manchester1994h,sykes2005observation,thurmer2022role}. 
Therefore, Pt(111) and Pd(111) surfaces provide representative case studies of adsorption and coupled bulk hydrogen absorption.

For a surface in equilibrium with a hydrogen reservoir, the particle number of hydrogen $N_H$ fluctuates and the appropriate description is the grand-canonical $\mu VT$ ensemble which constrains the chemical potential of hydrogen $\mu_H$ instead.
Conventional grand-canonical Monte Carlo (GCMC) samples this distribution through repeated insertion and deletion moves and typically requires simulations across a grid of chemical potentials. OMNIA instead conditions a single sampler on temperature $T$ and particle number $N_H$, learning the canonical distributions of hydrogen at different $N_H$. 
Thus, following Eq.~\ref{eq:ensemble_composition}, the path-weight estimates allow a single OMNIA model per metal substrate to recover the grand-canonical distribution and adsorption isotherm $\langle\theta_{\mathrm H}\rangle(T,\mu_{\mathrm H})$ across chemical potentials $\mu_H$ without retraining or resampling separately at each value (Methods).

\begin{figure}[!htbp]
  \centering
  \includegraphics[width=\textwidth]{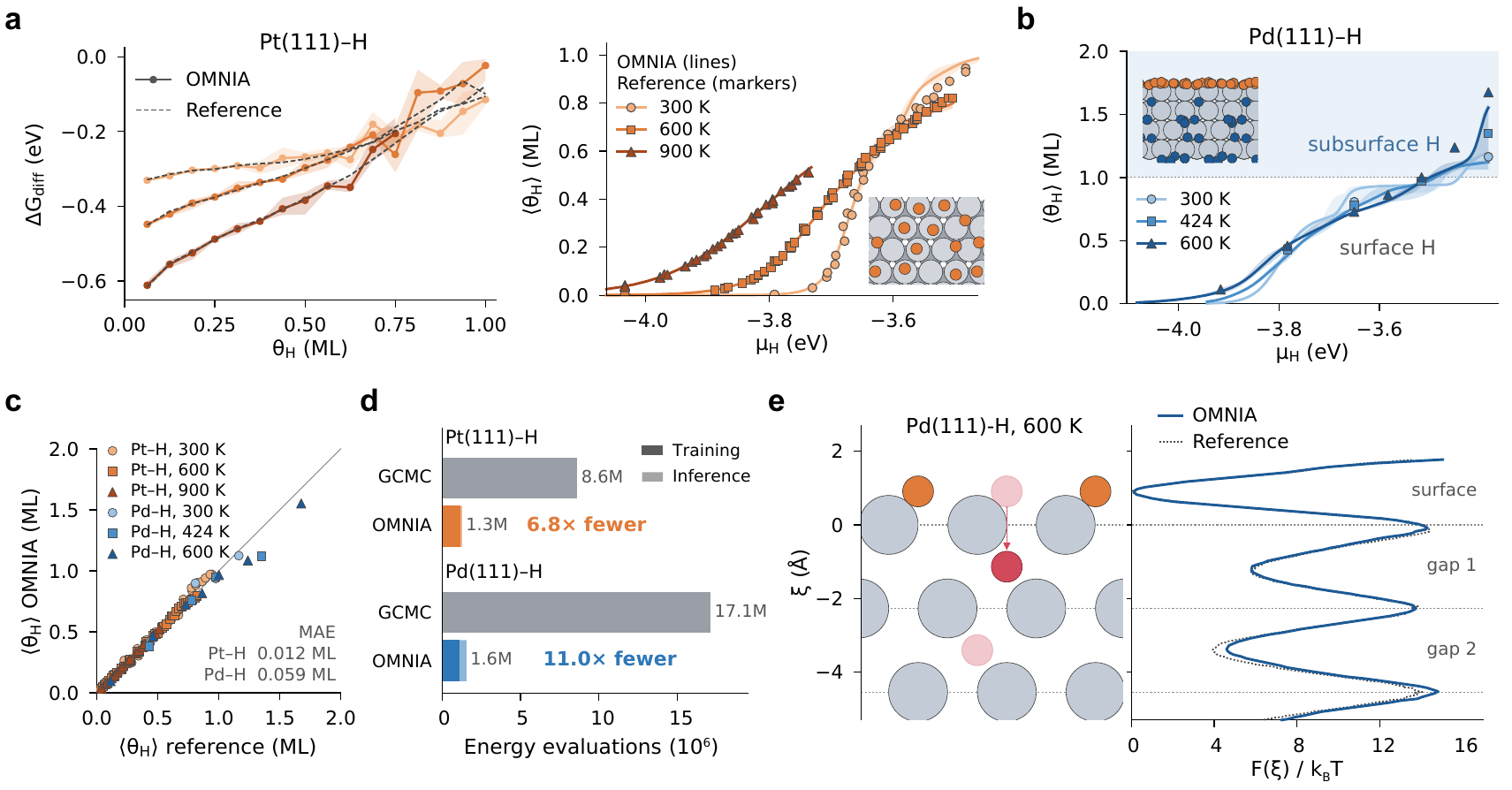}
\caption{\textbf{Grand-canonical hydrogen adsorption and absorption on metal surfaces.}
\textbf{a.} Differential adsorption free energies $\Delta G_\text{diff}$ (left) and adsorption isotherms (right) of hydrogen on Pt(111) at 300, 600 and 900\,K, comparing OMNIA with reference calculations, where $\theta_{\mathrm{H}}$ denotes hydrogen coverage in monolayers (ML), with 1 ML corresponding to one hydrogen atom per surface metal atom. The inset illustrates hydrogen adsorption on the Pt surface.
\textbf{b.} Hydrogen adsorption isotherms on Pd(111) at 300, 424 and 600\,K, showing subsurface hydrogen absorption beyond 1 ML coverage. The inset illustrates the Pd surface (orange) and subsurface (blue) hydrogen configurations. For \textbf{a, b}, shaded regions indicate standard deviation across random seeds.
\textbf{c.} Hydrogen coverages $\langle\theta_\mathrm{H}\rangle$ predicted by OMNIA against reference grand-canonical Monte Carlo (GCMC) results across temperatures and chemical potentials.
\textbf{d.} Computational costs measured by energy evaluations, including training and inference for OMNIA, compared to GCMC for Pt(111) and Pd(111), respectively.
\textbf{e.} Hydrogen penetration into Pd(111) at 600\,K and 1 ML coverage. The atomic illustration shows the penetration coordinate $\xi$, while the potential of mean force for one hydrogen atom predicted by OMNIA is compared with reference umbrella sampling.}
  \label{fig4}
\end{figure}

We apply this construction to hydrogen on Pt(111) and Pd(111), described by MACE-MH-1~\cite{batatia2025cross}, with the results shown in Fig.~\ref{fig4}. 
The reference data are obtained from parallel-tempered GCMC and weighted-histogram analysis~\cite{kumar1992weighted} across the studied temperature range (see Methods for details). 
As shown in Fig.~\ref{fig4}a, OMNIA reproduces the coverage-dependent differential adsorption free energy $\Delta G_\mathrm{diff}$ on Pt(111), defined as the free energy change associated with adding one hydrogen atom at a given coverage, together with the corresponding adsorption isotherms at multiple temperatures. 
The adsorption free energies agree with the reference to within approximately $0.01$-$0.02$\,eV, while the predicted coverages closely follow the reference calculations.
Pd(111) provides a more challenging scenario in which surface adsorption competes with subsurface absorption. In contrast to Pt(111), which approaches saturation near one monolayer (ML) coverage, Pd(111) continues to accommodate hydrogen beyond $\langle\theta_{\mathrm H}\rangle=1$ as additional atoms enter subsurface interstitial sites, as illustrated in Fig.~\ref{fig4}b.
OMNIA reproduces this adsorption-absorption crossover and the corresponding isotherms across the studied temperatures, with a mean coverage error of $0.059$\,ML relative to reference GCMC (Fig.~\ref{fig4}c). 
Moreover, thanks to the efficient amortization over multiple chemical potentials $\mu_H$, OMNIA requires approximately 11-fold fewer energy evaluations than the reference GCMC calculations across the sampled conditions (Fig.~\ref{fig4}d) to achieve the accuracy above.

Beyond sampling the trained thermodynamic ensemble, OMNIA can be efficiently steered at inference time toward additional observables or configurations. 
We demonstrate this by steering the Pd(111) sampler along the height $\xi$ of a hydrogen atom relative to the surface and reconstructing the potential of mean force (PMF) for hydrogen penetration from the surface into the substrate (Fig.~\ref{fig4}e).
The resulting free energy profile agrees with reference umbrella sampling to within $0.39\,k_\mathrm BT$ error, and correctly captures the successive barriers associated with entering the first subsurface layer and penetrating further into the substrate. 
Thus, a single OMNIA model can not only resolve competing adsorption and absorption on catalytic surfaces, but can also be utilized through inference-time steering to map free energy landscapes for interlayer and in-plane interfacial processes without retraining.

\subsection*{Sodium storage in amorphous carbon}
We finally apply OMNIA to sodium storage in hard carbon, a leading anode material for sodium-ion batteries and a low-cost alternative to graphite-based lithium-ion systems for large-scale energy storage~\cite{stevens2000high,hwang2017sodium}. 
Unlike crystalline graphite, hard carbon is non-graphitizing and contains disordered graphene-like layers together with defects and nanoscale pores.
Sodium storage can therefore involve several distinct environments, including adsorption at defects and pore surfaces, insertion between locally stacked carbon layers and clustering within nanopores~\cite{stevens2000high,front2026density}. The balance between these mechanisms depends strongly on the carbon microstructure and density, making atomic-scale structure itself an important design variable for sodium-storage capacity. 
Grand-canonical simulations with machine learning interatomic potentials can resolve these processes~\cite{front2026density}, but conventional GCMC must be repeated for every carbon framework and becomes particularly inefficient when insertion into compact, disordered atomic configurations has low acceptance.

\begin{figure}[!tp]
  \centering
  \includegraphics[width=\textwidth]{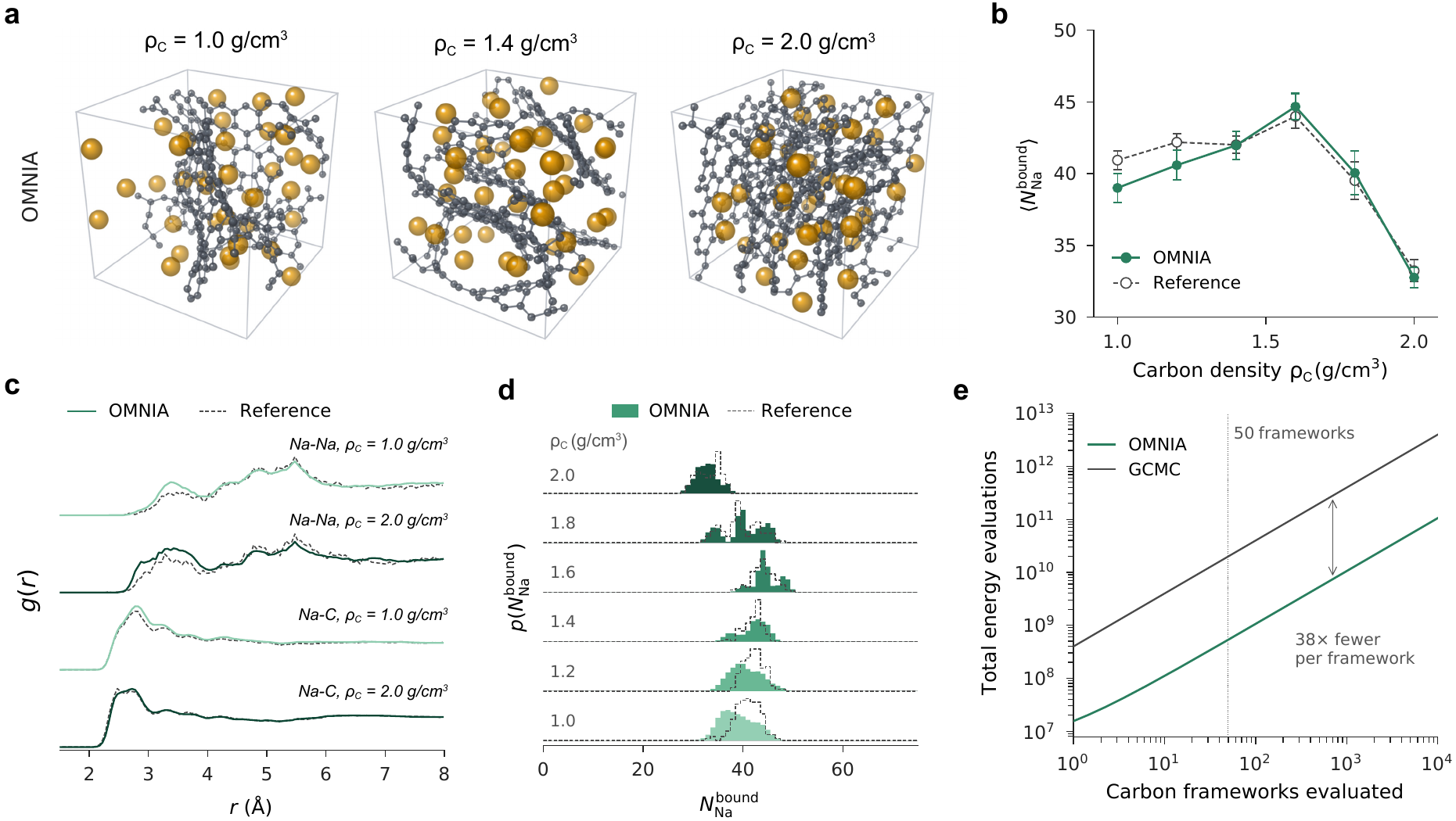}
 \caption{\textbf{Grand-canonical sodium storage in amorphous carbon.}
 OMNIA is trained on amorphous carbon frameworks spanning a range of carbon densities $\rho_\mathrm{C}$ at 300\,K and evaluated at a sodium chemical potential of $\mu_\mathrm{Na}=-1.24$\,eV. Reference results are obtained from grand-canonical Monte Carlo (GCMC). Only bound sodium atoms within 3.0\,\AA\ of carbon are considered in the analysis.
 \textbf{a.} Representative OMNIA-generated configurations at three carbon densities $\rho_\mathrm{C}=$1.0, 1.4 and 2.0\,g\,cm$^{-3}$.
 \textbf{b.} Mean number of bound sodium atoms $\langle N_\text{Na}^\text{bound}\rangle$ as a function of carbon density, compared with GCMC references. Each point is the grand-canonical mean of one framework, averaged over several carbon frameworks of that density. Error bars are the standard errors across these frameworks.
 \textbf{c.} Na-Na and Na-C radial distribution functions at carbon densities of 1.0 and 2.0\,g\,cm$^{-3}$, compared with GCMC references (dashed lines).
 \textbf{d.} Distributions of the bound sodium count $N_\text{Na}^\text{bound}$ at different carbon densities, pooled over the frameworks of each density and compared with GCMC references (dashed lines).
 \textbf{e.} Total number of energy evaluations as a function of the number of carbon frameworks, comparing the amortized cost of OMNIA with independent GCMC simulations.}
\label{fig5}
\end{figure}

We model hard carbon using amorphous carbon frameworks containing 250-500 carbon atoms at densities from 1.0 to 2.0\,g\,cm$^{-3}$, with sodium described by a C-Na Gaussian approximation potential~\cite{front2026density}. 
As for hydrogen on catalytic surfaces, OMNIA learns canonical distributions at fixed sodium count $N_\text{Na}$ and reconstructs the corresponding grand-canonical ensemble from their free energy. 
Here, however, a single model is additionally conditioned on the atomic structure of the carbon host, allowing the same sampler to generalize across thousands of distinct disordered frameworks rather than requiring an independent simulation for each one (Methods). We train on 8,032 frameworks spanning the full density range and evaluate the resulting model on representative carbon frameworks.
Some example hard carbon structures with stored sodium atoms at equilibrium at different carbon densities are shown in Fig.~\ref{fig5}a.

At ambient temperature ($T=300$\,K) and $\mu_\mathrm{Na}=-1.24$\,eV, corresponding to equilibrium with a metallic sodium reservoir, OMNIA reproduces the non-monotonic dependence of sodium storage on carbon density $\rho_C$ obtained from the GCMC reference (Fig.~\ref{fig5}b).
The number of sodium atoms bound to the carbon ($N_\text{Na}^\text{bound}$) increases from low density to a maximum near $\rho_C=$1.6\,g\,cm$^{-3}$ before decreasing again as the hard carbon framework becomes denser. 
This non-monotonic trend reflects a balance between available pore space and favorable Na-C binding. At high carbon density $\rho_C$, the confined pore and interlayer volume leaves insufficient space to accommodate sodium. At low $\rho_C$, larger pores provide more free volume, but a smaller fraction of sodium remains bound near the carbon framework and instead forms more weakly confined Na-rich clusters. 
Intermediate densities therefore balance sufficient free volume with abundant carbon-bound storage sites, maximizing sodium storage near the carbon framework and illustrating how grand-canonical sampling can guide hard-carbon structure design.

Beyond the average sodium loading, OMNIA also captures the underlying structural distributions, reproducing local Na-C and Na-Na radial distribution functions at both low and high carbon densities (Fig.~\ref{fig5}c) together with the bound-sodium populations across frameworks (Fig.~\ref{fig5}d). 
The largest deviations occur for structures with the lowest carbon density, where sodium redistribution within the large pores equilibrates more slowly even with GCMC.
As in the previous applications, OMNIA further benefits from amortization across thermodynamic and structural conditions. Here, a single model is trained across thousands of carbon frameworks, after which the structural samples for each new framework require approximately $38$-fold fewer GAP evaluations than direct GCMC (Fig.~\ref{fig5}e). 
Moreover, the one-time training cost is itself only about $1\%$ of the evaluations required for GCMC on a single framework. 
Together with inference-time steering, this capability of OMNIA opens opportunities for studying ion insertion and storage in porous materials~\cite{merlet2012molecular,sood2021electrochemical}, as well as molecular binding and docking in heterogeneous biological environments~\cite{kitchen2004docking,mobley2017predicting}.

\section*{Discussion}
In this work, we have introduced OMNIA as a universal neural sampling framework for equilibrium atomistic simulation. 
At the statistical mechanical level, different thermodynamic ensembles are expressed through conditional canonical distributions and the weights of variables released from fixed constraints, while competition between phases is resolved through their relative free energy learned from multiple expert models. 
At the modeling level, continuous and discrete diffusion processes describe atomic positions, cell degrees of freedom and chemical identities, while path weights provide statistical correction and estimates of partition functions and free energy. 
Using this generic construction, OMNIA accurately learns structural distributions and thermodynamic observables across $NVT$, $NPT$, $\Delta\mu NPT$ and $\mu VT$ ensembles, spanning oxide glasses, crystalline polymorphs, alloys, catalytic surfaces and porous carbon anodes. 
Across these applications, amortization over temperature, pressure, composition, particle number and host structure substantially reduces the number of energy evaluations required relative to conventional sampling. 
These results establish OMNIA as a general platform for equilibrium thermodynamics in atomistic simulation, applicable across diverse materials, ensembles and structural complexity.

Several challenges nevertheless remain. 
For example, although the systems studied here already span substantially different forms of structural and chemical disorder, extending the framework to extended defects such as grain boundaries and dislocations, reconstructed interfaces and electrochemical environments will require sampling much larger and more heterogeneous atomic configuration spaces across broader length scales.
Another direction is to condition a neural sampler directly on the thermodynamic ensemble itself, allowing a single pretrained model to adapt between different external constraints and reservoirs rather than relying on ensemble-specific constructions. 
More fundamentally, equilibrium sampling does not determine kinetics. Inference-time steering in OMNIA can reveal free energy landscapes and barriers along selected reaction coordinates, but physical transition rates still depend on the underlying dynamics. 
Combining neural samplers with rare event estimation methods such as committor learning represents a feasible route towards understanding the kinetics of atomistic systems~\cite{jung2023machine,tang2026breaking}.

More broadly, OMNIA addresses a fundamental challenge in atomistic simulation: while universal MLIPs such as MACE and UMA increasingly provide accurate and transferable energy and forces~\cite{batatia2025foundation,wood2026family}, thermodynamic sampling remains largely system specific and inefficient. 
At sufficient scale, a pretrained foundation OMNIA model could amortize sampling across materials, structures and thermodynamic conditions, replacing repeated trajectory generation with learned distributions that can be rapidly queried and steered on demand. Coupled with universal MLIPs, such foundation neural samplers could provide the missing pillar for scalable atomistic simulation, making complex phases, defects, disordered materials and heterogeneous interfaces accessible at unprecedented scale while connecting thermodynamic prediction to inverse design and materials discovery~\cite{cheng2026artificial,liu2025amorphous}.

\section*{Methods}

\subsection*{Diffusion samplers}
\paragraph{Generative process.} Each sampler is a diffusion model that transports a simple prior onto its
target ensemble~\cite{cheng2026atlas,blessing2026janus}. It generates the configuration $x$ from the reference ensemble jointly with the released
variables $z$, collected in $y=(x,z)$, so that its target is the full ensemble of Eq.~\eqref{eq:ensemble_composition},
\begin{equation}
p_\lambda(y)=\frac{\tilde p_\lambda(y)}{Z_\lambda},\qquad\tilde p_\lambda(y)=e^{-\beta\,[U(y)+R_\lambda(z)]},
\label{eq:target}
\end{equation}
where $U$ is the interatomic potential. Here, we consider the construction for continuous variables at a single thermodynamic state and describe chemical identities and the amortization over conditions below. The construction follows stochastic
interpolants\cite{JMLR:v26:23-1605}. A prior draw $y_0\sim\pi_0$ and a target draw $y_1\sim p_\lambda$ are connected by
the linear interpolant $y_t=y_0+\alpha(t)\,(y_1-y_0)$, with $\alpha$ increasing from $\alpha(0)=0$ to $\alpha(1)=1$,
whose marginals $p_t$ connect the prior to the target. The same marginals are realized by the stochastic differential
equation
\begin{equation}
\mathrm dy_t=\big[b_\theta(y_t,t)+\varepsilon(t)\,\nabla\varphi_\theta(y_t,t)\big]\,\mathrm dt
+\sqrt{2\varepsilon(t)}\,\mathrm dW_t,\qquad y_0\sim\pi_0,
\label{eq:sde}
\end{equation}
with velocity $b_\theta(y,t)\approx\mathbb E[\dot\alpha(t)\,(y_1-y_0)\mid y_t=y]$ and score $\nabla\log p_t(y)\approx \nabla\varphi_\theta(y,t)$. The Langevin strength
$\varepsilon(t)\ge0$ compensates the injected noise, so the marginals do not depend on
$\varepsilon$, and $\varepsilon=0$ recovers a deterministic flow. Parameterizing the energy $-\varphi_\theta/\beta$ and learning both, velocity and score, enables more sophisticated inference routines and allows for computing importance weights as discussed next.

\paragraph{Inference, path weights and free energy.} Configurations are generated by integrating
Eq.~\eqref{eq:sde} from a prior draw with the Euler-Maruyama scheme on the grid $t_k=kh$, $h=1/K$, so that every step
is drawn from the Gaussian transition kernel
\begin{equation}
\overrightarrow{p}_{\theta,k}(y_{k+1}\mid y_k)=\mathcal N\Big(y_{k+1};\,y_k+h\,\big[b_\theta+\varepsilon\,\nabla\varphi_\theta\big](y_k,t_k),\,2\varepsilon(t_k)\,h\,I\Big).
\label{eq:em}
\end{equation}
The time reversal of Eq.~\eqref{eq:sde}, whose drift subtracts rather than adds the score term~\cite{anderson1982reverse},
defines backward kernels $\overleftarrow{p}_{\theta,k}(y_k\mid y_{k+1})$ of the same form, with the drift
$b_\theta-\varepsilon\nabla\varphi_\theta$ evaluated at $(y_{k+1},t_{k+1})$. Each generated path then carries the log weight
\begin{equation}
\log w=\log\tilde p_\lambda(y_K)-\log\pi_0(y_0)+\sum_{k=0}^{K-1}\log\frac{\overleftarrow{p}_{\theta,k}(y_k\mid y_{k+1})}{\overrightarrow{p}_{\theta,k}(y_{k+1}\mid y_k)},
\label{eq:logw}
\end{equation}
which satisfies $\mathbb E[w]=Z_\lambda$ for any learned fields, step count and Langevin
strength~\cite{he2026free}; the quality of the model only affects the variance of $\log w$. Over $n$ independent paths, $\log\hat Z_\lambda=\operatorname{logsumexp}_i\log
w_i-\log n$ estimates the free energy $-k_\mathrm BT\log Z_\lambda$, and self-normalized averages
$\sum_iw_i\,O(y^{(i)})/\sum_iw_i$ reweight observables to the exact ensemble, at the cost of one energy evaluation per
path. 

Since the model also learns an energy, every intermediate marginal is available as an unnormalized density
$e^{\varphi_\theta(\cdot,t)}$, which permits sequential Monte Carlo (SMC)~\cite{delmoral2006sequential} along the
learned path. With intermediate targets $\log\gamma_0=\log\pi_0$, $\log\gamma_k=\varphi_\theta(\cdot,t_k)$
for $0<k<K$ and $\log\gamma_K=\log\tilde p_\lambda$, each of $n$ particles is propagated by
$\overrightarrow{p}_{\theta,k}$ and accumulates the incremental weights
\begin{equation}
\Delta\log w_k=\log\gamma_{k+1}(y_{k+1})-\log\gamma_k(y_k)+\log\frac{\overleftarrow{p}_{\theta,k}(y_k\mid y_{k+1})}{\overrightarrow{p}_{\theta,k}(y_{k+1}\mid y_k)},
\label{eq:smc}
\end{equation}
which sum to Eq.~\eqref{eq:logw} along a path. The population is then resampled whenever its effective sample size falls
below a certain threshold. Furthermore, optional Markov chain Monte Carlo (MCMC) moves that leave $\gamma_{k+1}$ invariant can be used after each step to move the particles towards the
intermediate target.

\paragraph{Data-free training.} Velocity and score are conditional expectations over the interpolant, so they can be
learned by least-squares regression\cite{JMLR:v26:23-1605}. For a prior draw $y_0$, a terminal
$y_1$ and a time $t$, the targets for $b_\theta$ and $\nabla\varphi_\theta$ at $(y_t,t)$ are
\begin{equation}
\hat b=\dot\alpha(t)\,(y_1-y_0),\qquad
\hat s=\nabla\log\pi_0(y_0)+\nabla\log p_\lambda(y_1),
\label{eq:targets}
\end{equation}
whose conditional expectations given $y_t$ are the exact velocity and score. Note that \eqref{eq:targets} uses the target score identity~\cite{de2024target} (Supplementary Section~\ref{si:score}) which contains the force at the target distribution as learning signal. The two fields are therefore fitted by the regression loss
\begin{equation}
\mathcal L_\mathrm{reg}(\theta)=\mathbb E\Big[\big\|b_\theta(y_t,t)-\hat b\big\|^2
+\big\|\nabla\varphi_\theta(y_t,t)-\hat s\big\|^2\Big],
\label{eq:loss_reg}
\end{equation}
with the expectation over times $t\sim\mathcal U[0,1]$, prior draws $y_0\sim\pi_0$ and terminals $y_1\sim p_\lambda$. Furthermore, the sampler is amortized over thermodynamic conditions~\cite{schebek2024efficient,cheng2026atlas} where the network takes the conditions $\lambda$ as additional inputs, e.g., $T$ and $P$ in $NPT$, and the expectation also runs over conditions
drawn from a window, so that one network represents the whole family of targets. Terminals $y_1\sim p_\lambda$,
however, are what a sampler is meant to produce and are not available a
priori. OMNIA replaces them by its own samples: each training round randomly draws conditions $\lambda$, generates
configurations with the current model, labels them once with forces, stores them with their conditions in a
replay buffer, and takes gradient steps on Eq.~\eqref{eq:loss_reg} (Supplementary Section~\ref{si:algorithms}). The family of targets is a fixed point of this iteration, i.e., if
the model samples every target exactly, the regression targets are those of the true interpolant and training is
stationary~\cite{havens2026flow,blessing2026bridge}.
We add an additional learning signal for the energy model $\varphi_\theta$ by regularizing the residual of the continuity equation,
\begin{equation}
r_\theta(y,t)=\partial_t\varphi_\theta(y,t)+b_\theta(y,t)\cdot\nabla\varphi_\theta(y,t)+\nabla\cdot b_\theta(y,t),
\label{eq:residual}
\end{equation}
by minimizing its variance at fixed time and conditions, together with the mismatch of $\varphi_\theta$ to $\log\pi_0$ and $\log\tilde p_\lambda$
at $t=0$ and $t=1$. The loss is evaluated on interpolant points between prior draws and buffered terminals, so it needs no simulation of the sampler (Supplementary Section~\ref{si:consistency}).

\subsection*{Canonical (NVT) ensembles}
\paragraph{Oxide glasses.} Multicomponent oxide glasses of the SiO$_2$-P$_2$O$_5$-Al$_2$O$_3$-Li$_2$O-CaO family are
described by the Bertani-Menziani-Pedone interatomic potentials~\cite{bertani2021improved}, with $N=255$ atoms in a fixed
periodic cubic cell of edge $L=14.95$\,\AA{} at $T=1200$\,K. The composition, which fixes the species of all sites, is
part of $\lambda$, and no variable is released, so that $z$ is empty and $y=x=\mathbf q\in[0,1)^{3N}$ are the fractional
coordinates of the atoms, sampled from
\begin{equation}
  p_\lambda(\mathbf q)\propto e^{-\beta\,U(\mathbf q)},
\end{equation}
where, here and below, $U$ is evaluated at the Cartesian positions defined by $y$, i.e., $L\mathbf q$. By the chain rule, the score of the target in fractional coordinates is the Cartesian force
$\mathbf F=-\nabla U$ scaled by the cell, $\nabla_{\mathbf q}\log p_\lambda=\beta L\,\mathbf F$. The prior is
uniform over the cell. The sampler is amortized over compositions drawn uniformly from the five-oxide simplex, so
that a single model covers many glasses. After every generation step, the samples are refined by one Hamiltonian
MC trajectory on the learned density $e^{\varphi_\theta}$, which requires no further energy evaluations. 
The reference is 32 independent melt-quench simulations per composition with the same interatomic potentials (Supplementary Section~\ref{si:oxides}).

\subsection*{Isothermal-isobaric (NPT) ensembles}
\paragraph{SiO$_2$ and Ti.} We compute the pressure-temperature phase diagrams of SiO$_2$, with the polymorphs
$\alpha$-quartz, coesite and stishovite, and of Ti, with its $\alpha$ (hcp), $\omega$ and $\beta$ (bcc) phases. SiO$_2$
is described by MACE-MH-1~\cite{batatia2025cross} where the sampler is amortized over a temperature and pressure window with 300-1800\,K and 0-12\,GPa, respectively, 
and Ti by MACE-MPA-0~\cite{batatia2025foundation} over 300-1600\,K and 0-20\,GPa. Each configuration contains
$N=216$ atoms in a fully flexible periodic cell. Here $x=\mathbf q\in[0,1)^{3N}$ are the fractional coordinates of the
atoms with the centre of mass fixed, and the released cell is $z=(v,\mathbf e)$, the logarithmic volume $v=\log V$ and
five deviatoric log-strains $\mathbf e=(e_1,\dots,e_5)$ that change the cell shape at fixed volume, so that
$y=(\mathbf q,v,\mathbf e)$. The cell matrix is
$H=e^{v/3}H_0\exp(\sum_ke_kB_k)$, where $H_0$ is the reference cell of the phase scaled to unit volume and $B_k$ is an
orthonormal basis of the traceless symmetric $3\times3$ matrices, so that $\det H=V$ and rigid rotations of the cell are
excluded. The target is
\begin{equation}
  p_\lambda(\mathbf q,v,\mathbf e)\propto e^{-\beta\,[U(\mathbf q,v,\mathbf e)+PV]+Nv},
\end{equation}
with $\lambda=(T,P)$, where $U$ is evaluated at the positions $\mathbf q_iH$ and $e^{Nv}$ is the Jacobian of the change
to fractional coordinates and logarithmic volume, with the volume measure $\mathrm dV/V$~\cite{frenkel2023understanding}. Each phase has its own
sampler, amortized over $T$ and $P$, whose prior is an Einstein crystal
around the lattice sites of the phase together with a Gaussian over the six cell coordinates, both fitted to short
MC runs with the same potential. Positions and cell coordinates are transported jointly by
Eq.~\eqref{eq:sde}. 
The Gibbs free energy $G_\alpha(T,P)=-k_\mathrm BT\log Z_{\lambda,\alpha}$ of each phase
$\alpha$ is estimated every 50\,K and 4\,GPa by SMC along the learned path, on a finer time grid than in training. At each
condition the stable phase is $\arg\max_\alpha p_\lambda(\alpha)=\arg\min_\alpha G_\alpha(T,P)$
(Eq.~\eqref{eq:phase_ensemble_composition}), and phase boundaries are located where the free energy of two
phases cross. The reference is non-reversible parallel tempering~\cite{syed2022nonreversible} with one ladder
per phase and pressure, analysed with MBAR~\cite{shirts2008statistically} (Supplementary Section~\ref{si:npt}).

\subsection*{Semi-grand-canonical ($\Delta \mu$NPT) ensembles}
\paragraph{Cu-Ag and Cu-Ni.} We compute the temperature-composition phase diagrams of Cu-Ag, a eutectic system,
and Cu-Ni, which melts through a narrow lens, both described by embedded-atom
potentials~\cite{williams2006embedded,fischer2019systematic}. Each configuration contains $N=108$ atoms in a periodic
cubic cell. Here $x=\mathbf q\in[0,1)^{3N}$ are the fractional coordinates with the centre of mass fixed, and the
released variables are $z=(\mathbf a,v)$, the species $\mathbf a\in\{\mathrm A,\mathrm B\}^N$ and the logarithmic
volume $v=\log V$, so that $y=(\mathbf q,\mathbf a,v)$ is drawn from
\begin{equation}
  p_\lambda(\mathbf q,\mathbf a,v)\propto
  e^{-\beta\,[U(\mathbf q,\mathbf a,v)+PV-\Delta\mu\,N_\mathrm B(\mathbf a)]+Nv}
\end{equation}
with $\lambda=(T,\Delta\mu)$ at $P=0$, where $N_\mathrm B$ counts the atoms of element B (Ag in Cu-Ag, Cu in Cu-Ni)
and $\Delta\mu=\mu_\mathrm B-\mu_\mathrm A$. Each phase has its own sampler,
amortized over $T$ and $\Delta\mu$: a Cu-rich fcc solid, an Ag-rich fcc solid and a liquid for Cu-Ag, and an fcc solid and a liquid for
Cu-Ni. The samplers use an Einstein crystal on the fcc lattice for the solids as prior distribution $\pi_0$ and uniformly distributed positions without a reference lattice for the liquids.
Positions and volume follow Eq.~\eqref{eq:sde}, while the species are generated by masked discrete diffusion, described next.

\paragraph{Chemical identities.} The species are transported by a masked discrete diffusion
model\cite{blessing2026janus}. Its interpolant hides the sites of a terminal configuration independently: given
$\mathbf a_1$, site $i$ shows its element $a_{1,i}$ with probability $\alpha_a(t)=t$ and an absorbing mask symbol
$\texttt M$ otherwise, so that $t=0$ is the fully masked state and no site ever shows a wrong element. Generation
reverses this process. Every site starts masked and is revealed exactly once, in a uniformly random order at the rate
set by $\alpha_a$, and keeps its element thereafter. The element is drawn from a learned denoising posterior
$q_{\theta,i}(k\mid y_t,t)$, the distribution of the site's terminal element given the current state $y_t$, i.e., the elements
revealed so far and the current positions and volume. 
Like the velocity and the score, the posterior is learned from the sampler's own terminal configurations, using a labeled cross-entropy loss:
\begin{equation}
  \mathcal L_\mathrm{species}=-\,\mathbb E\Big[\sum_{i:\,a_{t,i}=\texttt M}\ \sum_k p_{ik}\log
  q_{\theta,i}(k\mid y_t,t)\Big],\qquad
  p_{ik}=\frac{e^{\beta(\mu_k-\Delta U_{i\to k})}}{\sum_{k'}e^{\beta(\mu_{k'}-\Delta U_{i\to k'})}},
  \label{eq:species-loss}
\end{equation}
with labels $p_{ik}$ and $\Delta U_{i\to k}$ being the energy change of re-typing site $i$ of the terminal to element $k$ at fixed positions
and volume. Note that the Boltzmann distribution is again a fixed point of the training iteration\cite{blessing2026janus}. Furthermore,  the path weight of Eq.~\eqref{eq:logw} are adapted such that they incorporate the discrete part (Supplementary Section~\ref{si:diffusion}).

\paragraph{Phase boundaries.} The semi-grand potential of phase $\alpha$,
$\Phi_\alpha(T,\Delta\mu)=-k_\mathrm BT\log Z_{\lambda,\alpha}$ (Eq.~\eqref{eq:phase_ensemble_composition}), is
related to the mean composition of the phase by
$\partial\Phi_\alpha/\partial\Delta\mu=-\langle N_\mathrm B\rangle_\alpha$. The compositions that an expert generates on a
grid of $\Delta\mu$ along an isotherm therefore determine $\Phi_\alpha$ up to a constant, which is fixed at a pure element, whose absolute free energy in that phase is taken from the reference calculations. Two phases
coexist where their potentials cross, $\Phi_\alpha(T,\Delta\mu^\star)=\Phi_{\alpha'}(T,\Delta\mu^\star)$, and the
coexisting compositions are the mean compositions $\langle N_\mathrm B\rangle/N$ of the two experts at $\Delta\mu^\star$.
Repeating this construction at every temperature traces the phase boundaries. The reference is semi-grand-canonical
MC with the same potentials, complemented by fixed-composition ladders for the Cu-Ag solvus (Supplementary
Section~\ref{si:alloys}).

\subsection*{Grand canonical ($\mu$VT) ensembles}
\paragraph{Hydrogen on Pt(111) and Pd(111).} We compute the adsorption isotherms of hydrogen on Pt(111) and Pd(111),
described by MACE-MH-1~\cite{batatia2025cross}. The surfaces are $4\times4\times6$ slabs with 96 metal atoms,
periodic in the surface plane and held fixed, in equilibrium with a hydrogen reservoir at chemical potential
$\mu_\mathrm H$. Here $x=(\mathbf x_1,\dots,\mathbf x_N)\in D^N$ are the Cartesian positions of $N$ hydrogen atoms in a
domain $D$ around the surface, and the released variable is the particle number $z=N$, so that $y=(x,N)$ is drawn from
\begin{equation}
  p_\lambda(x,N)\propto\frac{1}{N!\,\Lambda^{3N}}\,e^{-\beta\,[U_N(x)-\mu_\mathrm HN]}
  \label{eq:gc-target}
\end{equation}
with $\lambda=(T,\mu_\mathrm H)$, where $U_N$ is the binding energy of the $N$ hydrogen atoms to the slab. The prefactor,
with $\Lambda$ the thermal wavelength of the atoms, accounts for their indistinguishability and their momenta. The
domain $D$ includes the first subsurface layer for Pt and four interlayer gaps for Pd.

Unlike the volume or the species, the particle number changes the dimension of $x$. OMNIA therefore does not generate
$N$ jointly with the positions, but learns the conditional distributions of Eq.~\eqref{eq:ensemble_composition},
$p_\lambda(x\mid N)\propto e^{-\beta U_N(x)}$, with one sampler per metal that is conditioned on $N$ and amortized over
$T$ (Supplementary Section~\ref{si:conditional}). The path weights of Eq.~\eqref{eq:logw} then estimate the canonical
partition functions $Z_N(T)=\int_{D^N}e^{-\beta U_N(x)}\,\mathrm dx$, which give the weight of every particle number,
\begin{equation}
  p_\lambda(N)\propto\frac{e^{\beta\mu_\mathrm HN}}{N!\,\Lambda^{3N}}\,Z_N(T).
  \label{eq:gc-weight}
\end{equation}
Since $\mu_\mathrm H$ enters only through this prefactor, the estimates of $Z_N$ at one temperature yield the isotherm
at any chemical potential without further sampling. To give every $N$ the same state space, the sampler acts on a fixed
number $M$ of slots: $N$ of them carry hydrogen, and the remaining $M-N$ carry non-interacting ghost atoms that follow a
fixed normalized density $g$ on $D$ and therefore leave $Z_N$ unchanged (Supplementary Section~\ref{si:slots}). The
density $g$, which also serves as the prior $\pi_0$ of every slot, places narrow Gaussians on the binding sites of a
single hydrogen atom over a uniform background and is computed once from the potential, without reference data
(Supplementary Section~\ref{si:site-prior}). The sampler for Pt covers 300-900\,K and up to 20 hydrogen atoms, and the
sampler for Pd 300-600\,K and up to 32 atoms. The reference is GCMC with the same
potential (Supplementary Section~\ref{si:surfaces}).

\paragraph{Potential of mean force for absorption.} We additionally compute the potential of mean force $F(\xi)$ of
one hydrogen atom along its height $\xi$ above the top plane of Pd(111), at 600\,K and $N=16$, which is a coverage of
one monolayer on the 16 surface sites of the $4\times4$ cell. Since the hydrogen atoms are indistinguishable, we
restrain one tagged atom, which gives the profile of any one of them. The tagged atom is restrained by
harmonic umbrella windows~\cite{torrie1977nonphysical}, $r_k=-\tfrac12\beta\kappa\,(\xi-\xi_{0,k})^2$, with
$\kappa=3$\,eV\,\AA$^{-2}$ and window centres $\xi_{0,k}$ from $-5$ to 2\,\AA{} in steps of 0.2\,\AA. Each window is the tilted ensemble
$\propto p_\lambda(x\mid N)\,e^{r_k(x)}$, which we sample by inference-time steering of the trained Pd-H sampler, without
retraining~\cite{cheng2026atlas,he2026rne,blessing2026janus}: the SMC of Eq.~\eqref{eq:smc} runs with reward-tilted
intermediate targets and a proposal guided by the gradient of the reward (Supplementary Section~\ref{si:steering}). The
path weights give the partition function of every window, so that the reweighted height histograms $\rho(\xi)$ of all
windows combine into $F(\xi)=-k_\mathrm BT\log\rho(\xi)$ without fitted offsets. The reference is umbrella sampling with
the same potential (Supplementary Section~\ref{si:pmf}).

\paragraph{Sodium storage in amorphous carbon.} We compute the sodium uptake of amorphous carbon frameworks at
$T=300$\,K and a sodium chemical potential of $\mu_\mathrm{Na}=-1.24$\,eV, which corresponds to equilibrium with metallic sodium, described by the C-Na Gaussian approximation
potential (GAP) of Front et al.~\cite{front2026density}. Each framework $F_\mathrm C$ is generated by melt-quenching, contains
$N_\mathrm C=250$ to 500 carbon atoms in a periodic cubic cell of edge $L=17.08$\,\AA{} and is held fixed. Here $x=(\mathbf x_1,\dots,\mathbf x_N)\in[0,L)^{3N}$ are the positions of $N$ sodium
atoms and $z=N$, so that $y=(x,N)$ is drawn from the grand-canonical ensemble of Eq.~\eqref{eq:gc-target} with
$\lambda=(T,\mu_\mathrm{Na},F_\mathrm C)$, where $U_N(x;F_\mathrm C)$ is now the binding energy of the sodium atoms to
the framework. The frameworks are independent draws of the melt-quench, so many different frameworks share one carbon density.
OMNIA samples the sodium in each framework separately, conditioned on the positions of its carbon atoms. Results at a
given carbon density are averages over the frameworks of that density and, within each framework, over the
grand-canonical ensemble of the sodium.

As for hydrogen, OMNIA learns the conditional distributions
$p_\lambda(x\mid N)\propto e^{-\beta U_N(x;F_\mathrm C)}$ and obtains the weights $p_\lambda(N)$ from
Eq.~\eqref{eq:gc-weight}. Normalized over $N$, $p_\lambda(N)$ is the
probability of finding $N$ sodium atoms in the framework at the chemical potential $\mu_\mathrm{Na}$. A
grand-canonical sample thus consists of two draws: $N$ from $p_\lambda(N)$, then $x$ from $p_\lambda(x\mid N)$. A single sampler is conditioned on $N\le200$ and amortized over all frameworks: the network
encodes the carbon atoms once per framework and transports only the sodium positions. The prior is a lattice gas on the
binding sites of the framework, which are located with a C-Na pair potential fitted once to the GAP, so that the prior
requires no GAP evaluation for a new framework. For each framework, $Z_N$ is
estimated at $N=1,10,\dots,80$ by SMC along the learned path, and the weights of all other $N$ follow from a smooth fit.
Structural averages are taken from samples at the most probable $N$ and count only the bound sodium atoms,
those within 3.0\,\AA{} of a carbon atom; their number is $N_\text{Na}^\text{bound}$. Its distribution in one
framework is
\begin{equation}
  p_\lambda(N_\text{Na}^\text{bound})=\sum_Np_\lambda(N)\,p_\lambda(N_\text{Na}^\text{bound}\mid N),
  \label{eq:bound-sodium}
\end{equation}
where $p_\lambda(N_\text{Na}^\text{bound}\mid N)$ is the distribution of the bound count at fixed $N$. The reference is GCMC with replica exchange in $\mu_\mathrm{Na}$ on 50 of the frameworks (Supplementary
Section~\ref{si:sodium}).

\section*{Acknowledgments}
The authors thank Jutta Rogal for helpful discussions. JH acknowledges support from the University of Cambridge Harding Distinguished Postgraduate Scholars Programme.

\newpage

\begingroup
\sffamily
\raggedright
\setlength{\parindent}{0pt}

\vspace*{-24pt}

{\bfseries\fontsize{18}{22}\selectfont
OMNIA: Universal Neural Samplers for Atomistic Simulation: Supplementary Information
\par}
\vspace{9pt}
{\bfseries\selectfont
Mouyang Cheng$^{1,2,3,*,\#}$, Denis Blessing$^{1,4,*,\#}$, Jiajun He$^{1,5,\#}$, Mingda Li$^{2,6}$, Carles Domingo-Enrich$^{1,\dagger}$, and Yuanqi Du$^{1,\dagger}$
\par}
\vspace{12pt}
{\fontsize{9.5}{11.5}\selectfont
$^{1}$Microsoft Research New England, Cambridge, MA 02142, USA\\
$^{2}$Center for Computational Science and Engineering, MIT, Cambridge, MA 02139, USA\\
$^{3}$Department of Materials Science and Engineering, MIT, Cambridge, MA 02139, USA\\
$^{4}$Karlsruhe Institute of Technology, 76131 Karlsruhe, Germany\\
$^{5}$University of Cambridge, Cambridge, CB2 1PZ, UK\\
$^{6}$Department of Nuclear Science and Engineering, MIT, Cambridge, MA 02139, USA\\
$^{*}$These authors contributed equally.\\
$^{\#}$Work done during internship at Microsoft Research.\\
$^\dagger$Correspondence. Email: carlesd@microsoft.com, yuanqidu@microsoft.com
\par}
\endgroup

\addtocontents{toc}{\protect\setcounter{tocdepth}{2}}
\tableofcontents
\setcounter{section}{0}
\setcounter{figure}{0}
\setcounter{table}{0}
\setcounter{equation}{0}
\setcounter{algorithm}{0}
\renewcommand{\thesection}{S\arabic{section}}
\renewcommand{\thefigure}{S\arabic{figure}}
\renewcommand{\thetable}{S\arabic{table}}
\renewcommand{\theequation}{S\arabic{equation}}
\renewcommand{\thealgorithm}{S\arabic{algorithm}}
\renewcommand{\theHsection}{Supp\arabic{section}}
\renewcommand{\theHfigure}{Supp\arabic{figure}}
\renewcommand{\theHtable}{Supp\arabic{table}}
\renewcommand{\theHequation}{Supp\arabic{equation}}
\renewcommand{\theHalgorithm}{Supp\arabic{algorithm}}

\section{Method details}
\label{si:framework}

This section collects the parts of OMNIA that are shared by all systems: the conditional samplers, the diffusion
process and its regression targets, the permutation coupling of the endpoints, the simulation-free consistency loss,
pseudocode for training and for inference by sequential Monte Carlo, and inference-time steering. The system-specific choices are given in
Sections~\ref{si:oxides}-\ref{si:sodium}.

\subsection{Conditional samplers}
\label{si:conditional}
For $\mu VT$, OMNIA learns the reference ensembles $p_\lambda(x\mid z)$ of Eq.~\eqref{eq:ensemble_composition} instead
of $p_\lambda(x,z)$, which simplifies the learning problem. The sampler generates $x$ at fixed $z$ and receives $z$ as
an input, like $\lambda$. Its target is
\begin{equation}
  p_\lambda(x\mid z)=\frac{\tilde p_\lambda(x\mid z)}{Z(z;\lambda)},\qquad\tilde p_\lambda(x\mid z)=e^{-\beta\,U(x,z)},
\end{equation}
since the reservoir term $R_\lambda(z)$ is constant at fixed $z$; for the slots of Section~\ref{si:slots},
$\tilde p_\lambda$ also carries the ghost factor. A prior draw $x_0\sim\pi_0(\cdot\mid z)$ is transported by
\begin{equation}
  \mathrm dx_t=\big[b_\theta(x_t,z,t)+\varepsilon(t)\,\nabla\varphi_\theta(x_t,z,t)\big]\,\mathrm dt
  +\sqrt{2\varepsilon(t)}\,\mathrm dW_t,\qquad x_0\sim\pi_0(\cdot\mid z),
\end{equation}
where $\nabla$ acts on $x$ only. The fields are fitted as in the Methods, on the interpolant
$x_t=\mathrm{wrap}(x_0+\alpha(t)\,\Delta)$, $\Delta=\mathrm{wrap}(x_1-x_0)$, of Section~\ref{si:diffusion} with
terminals $x_1\sim p_\lambda(\cdot\mid z)$ and the targets $\hat b=\dot\alpha(t)\,\Delta$ and $\hat s=\nabla\log\pi_0(x_0\mid z)+\nabla\log p_\lambda(x_1\mid z)$. The path
weight
\begin{equation}
  \log w=\log\tilde p_\lambda(x_K\mid z)-\log\pi_0(x_0\mid z)+\sum_{k=0}^{K-1}\log\frac{\overleftarrow{p}_{\theta,k}(x_k\mid x_{k+1},z)}{\overrightarrow{p}_{\theta,k}(x_{k+1}\mid x_k,z)}
\end{equation}
then satisfies $\mathbb E[w]=Z(z;\lambda)$, from which the weights $p_\lambda(z)$ follow.

\subsection{Diffusion process}
\label{si:diffusion}

\paragraph{Continuous channels.} A prior draw $y_0\sim\pi_0$ and a terminal $y_1$ are joined by the interpolant
\begin{equation}
  y_t=\mathrm{wrap}\big(y_0+\alpha(t)\,\Delta\big),\qquad\Delta=\mathrm{wrap}(y_1-y_0),
  \label{eq:si-interpolant}
\end{equation}
where $\alpha$ increases from $\alpha(0)=0$ to $\alpha(1)=1$. On periodic coordinates, $\mathrm{wrap}$ folds a
displacement back into the periodic cell, so that $\Delta$ is the minimum-image displacement between the two
endpoints~\cite{cheng2026atlas}. The oxide glasses, SiO$_2$, Ti and the alloys use fractional coordinates, for which
$\mathrm{wrap}(q)=q-\lfloor q+\tfrac12\rfloor\in[-\tfrac12,\tfrac12)$ on every component. Hydrogen and sodium use
Cartesian positions, for which $\mathrm{wrap}$ subtracts the corresponding lattice vectors of the cell. On non-periodic
coordinates, such as the cell coordinates, $\mathrm{wrap}$ is the identity.

Samples are generated by the Euler-Maruyama discretization of the stochastic differential equation
of the main text on $K$ uniform steps, $h=1/K$, with the forward kernels
\begin{equation}
  \overrightarrow{p}_{\theta,k}(y_{k+1}\mid y_k)=\mathcal N\big(y_{k+1};\,y_k+h\,[b_\theta+\varepsilon\nabla\varphi_\theta](y_k,t_k),\,2\varepsilon(t_k)h\,I\big).
\end{equation}
The backward kernels of the path weights are Gaussians with the time-reversed drift evaluated at the later point and the
same variance,
\begin{equation}
  \overleftarrow{p}_{\theta,k}(y_k\mid y_{k+1})=\mathcal N\big(y_k;\,y_{k+1}-h\,[b_\theta-\varepsilon\nabla\varphi_\theta](y_{k+1},t_{k+1}),\,2\varepsilon(t_k)h\,I\big).
\end{equation}
Because the score term compensates the injected noise, the Langevin strength $\varepsilon(t)$ can be chosen freely at
inference. Its upper limit is the stability of the Euler step in the stiffest wells, $\varepsilon h\beta k\lesssim2$ for a
well of curvature $k$.

\paragraph{Discrete channel.} Chemical identities follow the absorbing-mask process of the main text: given
$\mathbf a_1$, each site shows its element with probability $\alpha_a(t)=t$ and the mask $\texttt M$ otherwise.
Generation reveals the sites in a uniformly random order, $\mathrm{round}(t_kN)$ of them by step $k$, and draws each
element from the learned posterior $q_{\theta,i}$. With $K\ge N$ at most one site is revealed per network call, so that the
sampler is an any-order autoregressive model~\cite{blessing2026janus}. The realized log-probabilities of all reveals,
$\log q_\mathrm{disc}$, enter the path weight alongside the Gaussian kernels of the continuous channels.

\subsection{Regression targets}
\label{si:score}

Let the endpoints $(y_0,y_1)$ have joint density $q$, and let $p_t$ be the law of the
interpolant $y_t=\mathrm{wrap}\big(y_0+\alpha\,\mathrm{wrap}(y_1-y_0)\big)$ of
Eq.~\eqref{eq:si-interpolant}, $\alpha=\alpha(t)$. Then, on the torus and for any $\alpha$,
\begin{equation}
  \nabla\log p_t(y)=\mathbb E\big[(\nabla_{y_0}+\nabla_{y_1})\log q(y_0,y_1)\,\big|\,y_t=y\big].
  \label{eq:si-translation-identity}
\end{equation}
Indeed, translating both endpoints by the same $h$ leaves $\mathrm{wrap}(y_1-y_0)$ unchanged
and so translates $y_t$ by $h$. Hence, for any smooth test function $f$,
$\int f(y)\,p_t(y-h)\,\mathrm dy=\mathbb E[f(y_t+h)]=\int f(y_t)\,q(y_0-h,y_1-h)\,\mathrm dy_0\,\mathrm dy_1$,
and differentiating at $h=0$ gives Eq.~\eqref{eq:si-translation-identity}. For independent
endpoints, $q=\pi_0\otimes p_\lambda$, this yields the target
\begin{equation}
  \hat s=\nabla\log\pi_0(y_0)+\nabla\log p_\lambda(y_1),\qquad
  \mathbb E[\hat s\mid y_t=y]=\nabla\log p_t(y),
  \label{eq:si-score-target}
\end{equation}
which is independent of the schedule and has bounded variance at both ends of the interval.
For a uniform prior it reduces to the force at the terminal in units of $k_\mathrm BT$, the
target of Cheng et al.~\cite{cheng2026atlas}; Eq.~\eqref{eq:si-translation-identity} extends it to
non-uniform priors, such as the Einstein crystals, and to coupled endpoints
(Section~\ref{si:coupling}). The single-endpoint identities of Ref.~\cite{de2024target}, of which
Eq.~\eqref{eq:si-score-target} is a convex combination on $\mathbb R^n$, move one endpoint
with the other fixed. On periodic coordinates this fails: $\Delta$ jumps by a cell vector
when the endpoints are half a cell apart, which adds a surface term to $\nabla\log p_t$ that
these estimators miss, whereas moving both endpoints together never changes $\Delta$. On
non-periodic channels they hold; for the logarithmic volume of the alloys we use the target
$[(1-\alpha)\nabla\log\pi_0(y_0)+\alpha\nabla\log p_\lambda(y_1)]/D$, $D=(1-\alpha)^2+\alpha^2$
(Section~\ref{si:alloys}).

For the atomic coordinates, $\nabla\log p_\lambda(y_1)=-\beta\nabla U(y_1)$ is the force at the
terminal in units of $k_\mathrm BT$, so each buffered configuration needs one force
evaluation. The velocity target is $\hat b=\dot\alpha\,\Delta$, whose conditional expectation
is the probability-flow velocity of the interpolant. Both targets are evaluated on fresh prior
draws at every inner step, so the buffered labels are reused without storing prior samples.

\subsection{Permutation-coupled endpoints}
\label{si:coupling}

Neural samplers are usually trained on independent
endpoints~\cite{blessing2026bridge,havens2026flow,cheng2026atlas,blessing2026janus}. Same-species
particles of the prior and terminal samples are then paired at random, and the velocity target
averages over many crossing paths, which makes it noisy. Optimal transport couplings shorten
the paths~\cite{pooladian2023multisample,tong2024improving,klein2023equivariant}, but the joint distribution of $(y_0,y_1)$ under a
deterministic assignment has no smooth density, so the score target~\eqref{eq:si-score-target}
cannot be corrected for it. We instead draw a random permutation that favors nearby pairs, at
$O(N^2)$ cost. Because it
is random, the score target can be corrected for it exactly, with a closed-form term.

\paragraph{Kernel.} The kernel pairs each particle of the terminal $y_1$ with a particle of the
same species in a prior draw $u\sim\pi_0$, favoring nearby pairs. The terminal particles are
visited in a uniformly random order $\sigma$, drawn independently of the endpoints. Labeling
them $i=1,\dots,N$ in that order, particle $i$ takes an unused prior particle $p_i=j\in R_i$
with probability
\begin{equation} \label{eq:si-weights-permutation}
  w_{ij}=\frac{e^{-|r_{ij}|^2/\epsilon_m}}{Z_i},\qquad
  Z_i=\sum_{j\in R_i}e^{-|r_{ij}|^2/\epsilon_m},\qquad
  r_{ij}=\mathrm{wrap}(\mathbf y_{1,i}-\mathbf u_j),
\end{equation}
where $R_i$ is the set of still-unused prior particles of the same species as $i$,
$\mathbf y_{1,i}$ and $\mathbf u_j$ are the fractional positions of the two particles
(Cartesian for SiO$_2$, Ti and sodium), and the temperature $\epsilon_m$ sets how strongly
nearby pairs are favored. Reordering $u$ so that its $i$-th particle is the one paired with
terminal particle $i$ gives the prior endpoint of the interpolant, $y_0=Pu$, where $P$ is the
permutation matrix with $P_{i,p_i}=1$, acting on whole particles. Because particles of one species are exchangeable in both the target and the prior, and the
visiting order is random, this reordering leaves both endpoint laws unchanged. The prior draw is permuted before $\Delta$ is
computed, so that $y_{t=1}=y_1$ exactly, and every loss that builds interpolant points,
including the consistency loss (Section~\ref{si:consistency}), uses the same permutation.

\paragraph{Score target.} Given $(u,y_1)$, the pair $(\sigma,P)$ has probability
$K(\sigma,P\mid u,y_1)=K(P\mid u,y_1,\sigma)/N!$ with $K(P\mid u,y_1,\sigma)=\prod_iw_{i,p_i}$.
Since $\pi_0(u)=\pi_0(y_0)$ by exchangeability, the joint density is
\begin{equation}
  q(y_0,y_1,\sigma,P)=\pi_0(y_0)\,p_\lambda(y_1)\,K\big(\sigma,P\mid P^{-1}y_0,y_1\big).
  \label{eq:si-coupled-law}
\end{equation}
The proof of Eq.~\eqref{eq:si-translation-identity} translates the endpoints at fixed
$(\sigma,P)$, so it applies to this $q$. The factors $\pi_0(y_0)\,p_\lambda(y_1)$ of $q$ give the
target $\hat s$ of Eq.~\eqref{eq:si-score-target}, evaluated at $(y_0,y_1)$ with $y_0=Pu$, and
the kernel adds
\begin{equation}
  c=(\nabla_{y_0}+\nabla_{y_1})\log K\big(\sigma,P\mid P^{-1}y_0,y_1\big)
   =D_P\log K(P\mid u,y_1,\sigma),\qquad D_P=\nabla_{y_1}+P\,\nabla_u,
  \label{eq:si-coupling-correction}
\end{equation}
so that the score target becomes $\hat s+c$. The velocity target needs no correction, since
$\mathbb E[\dot\alpha\Delta\mid y_t]$ is the interpolant velocity for any coupling. Exactness
requires that $K$ be positive and differentiable in the endpoints at fixed $(\sigma,P)$, and
that $\sigma$ not depend on the endpoints. A deterministic assignment, such as the Hungarian
minimum-cost matching, violates the first condition.

\paragraph{Closed form.} For the weights of Eq.~\eqref{eq:si-weights-permutation},
$\log K=\sum_i\big[-|r_{i,p_i}|^2/\epsilon_m-\log Z_i\big]$. The $k$-th component of $D_P$,
$\nabla_{\mathbf y_{1,k}}+\nabla_{\mathbf u_{p_k}}$, translates the matched pair together and
leaves $r_{k,p_k}$ unchanged, so only the normalizers contribute: $Z_k$ through
$\mathbf y_{1,k}$, and every $Z_i$ with $i\le k$ through $\mathbf u_{p_k}$, which was still
available at those visits. With
$\nabla_{\mathbf y_{1,i}}|r_{ij}|^2=-\nabla_{\mathbf u_j}|r_{ij}|^2=2r_{ij}$,
\begin{equation}
  c_k=\frac{2}{\epsilon_m}\Big[\sum_{j\in R_k}w_{kj}\,r_{kj}-\sum_{i\le k}w_{i,p_k}\,r_{i,p_k}\Big]
  \label{eq:si-coupling-closed}
\end{equation}
in the coordinates in which $r_{ij}$ is measured; where these are Cartesian, $c$ is converted
to the coordinates of the force label. Since $|r_{ij}|^2$ is continuous across the cut of the
minimum image, no surface term arises. Translating all particles together leaves every
$r_{ij}$ unchanged, so $\sum_kc_k=0$. The correction enters the force label as $k_\mathrm BT\,c$.

\subsection{Simulation-free consistency loss}
\label{si:consistency}
The regression loss fits the velocity $b_\theta$ and the score $\nabla\varphi_\theta$ separately and determines
$\varphi_\theta$ only through its gradient. Sequential Monte Carlo, however, uses $e^{\varphi_\theta(\cdot,t)}$ as its
intermediate targets, so $\varphi_\theta$ itself should be the log-density that $b_\theta$ transports from the prior to
the target. This holds if $e^{\varphi_\theta}$ satisfies the continuity equation
$\partial_tp_t+\nabla\cdot(p_tb_\theta)=0$ up to its normalization, that is, if the residual $r_\theta(y,t)$ of
Eq.~\eqref{eq:residual} does not depend on $y$, and if $\varphi_\theta$ agrees with $\log\pi_0$ at $t=0$ and with
$\log\tilde p_\lambda$ at $t=1$ up to constants. Since $e^{\varphi_\theta}$ is not normalized, each of these three
quantities only has to be constant over configurations, not zero. The consistency loss therefore penalizes their
variances over configurations, and not their squares,
\begin{equation}
  \mathcal L_\mathrm{cons}=w_\mathrm{local}\,\mathbb E_t\operatorname{Var}\big[r_\theta(y_t,t)\big]
  +w_\mathrm{end}\Big(\operatorname{Var}\big[\varphi_\theta(y_0,0)-\log\pi_0(y_0)\big]
  +\operatorname{Var}\big[\varphi_\theta(y_1,1)-\log\tilde p_\lambda(y_1)\big]\Big),
  \label{eq:si-cons}
\end{equation}
where $y_t$ is the interpolant between a fresh prior draw $y_0$ and a buffered terminal $y_1$, as in the regression
loss. Each variance is taken over a group of configurations that share the time $t$ and the conditions $\lambda$,
because the constants depend on both. The loss needs no simulation of the sampler and no energy evaluation, since
$\log\tilde p_\lambda(y_1)$ is stored in the buffer.

The divergence $\nabla\cdot b_\theta$ in $r_\theta$ is estimated with Hutchinson's
estimator~\cite{hutchinson1989stochastic}. The variance of such a noisy estimate would also contain the noise of the
estimator. We therefore evaluate the residual twice with independent noise and use the covariance of the two estimates
within a group, which is an unbiased estimate of the variance of $r_\theta$. The gradient of $\mathcal L_\mathrm{cons}$ is
rescaled to the norm of the regression gradient before it is applied, so that neither loss dominates the update.

\subsection{Training and inference algorithms}
\label{si:algorithms}

Algorithm~\ref{alg:train} summarizes the training and Algorithm~\ref{alg:smc} the inference by sequential Monte Carlo.
In training, the interatomic potential is evaluated only when the generated configurations are labelled. The labels
are stored in the replay buffer and reused by many gradient steps. The consistency loss of
Section~\ref{si:consistency} is optional and needs no further evaluation of the potential.

\begin{algorithm}[t]
\caption{Training of an OMNIA sampler.}
\label{alg:train}
\begin{algorithmic}[1]
\Require interatomic potential; window of conditions $\lambda$; prior $\pi_0$; schedule $\alpha(t)$; Langevin strength
$\varepsilon$
\State initialize the network $\theta$ and an empty replay buffer $\mathcal B$
\For{each outer iteration}
  \State \textit{generate:} draw conditions $\lambda$ from the window and generate configurations $y_1$ with the current model (Eq.~\eqref{eq:em})
  \State \textit{label:} evaluate the potential at every $y_1$, which gives the energy and the forces, and for alloys the labels $p_{ik}$ of Eq.~\eqref{eq:species-loss}
  \State store every $y_1$ with its labels and conditions in $\mathcal B$, replacing the oldest entries
  \For{each gradient step}
    \State draw a mini-batch of $y_1$ with labels and conditions from $\mathcal B$, prior draws $y_0\sim\pi_0$ and times $t\sim\mathcal U[0,1]$
    \State \textit{couple} (optional): permute the atoms of $y_0$ to match $y_1$ (Section~\ref{si:coupling})
    \State \textit{interpolate:} form $y_t$ from $y_0$ and $y_1$ (Eq.~\eqref{eq:si-interpolant}); for alloys, mask each species with probability $1-t$
    \State \textit{regress:} gradient step on $\mathcal L_\mathrm{reg}$ (Eq.~\eqref{eq:loss_reg}), for alloys with $\mathcal L_\mathrm{species}$ added
    \State \textit{consistency} (optional): gradient step on $\mathcal L_\mathrm{cons}$ (Eq.~\eqref{eq:si-cons})
  \EndFor
\EndFor
\end{algorithmic}
\end{algorithm}

\begin{algorithm}[t]
\caption{Inference by sequential Monte Carlo at one condition $\lambda$.}
\label{alg:smc}
\begin{algorithmic}[1]
\Require trained network; number of particles $n$; number of steps $K$; intermediate targets $\gamma_k$ with
$\gamma_0=\pi_0$ and $\gamma_K=\tilde p_\lambda$
\State draw $y_0^{(i)}\sim\pi_0$ and set $\log w^{(i)}\gets0$ for $i=1,\dots,n$; set $\log\hat Z\gets0$
\For{$k=0,\dots,K-1$}
  \State \textit{propagate:} draw $y_{k+1}^{(i)}\sim\overrightarrow{p}_{\theta,k}(\cdot\mid y_k^{(i)})$ (Eq.~\eqref{eq:em})
  \State \textit{reweight:} $\log w^{(i)}\gets\log w^{(i)}+\Delta\log w_k^{(i)}$ (Eq.~\eqref{eq:smc})
  \If{the effective sample size $\big(\sum_iw^{(i)}\big)^2/\sum_i\big(w^{(i)}\big)^2$ is below $n/2$}
    \State $\log\hat Z\gets\log\hat Z+\operatorname{logsumexp}_i\log w^{(i)}-\log n$
    \State \textit{resample:} draw $n$ particles with probabilities $\propto w^{(i)}$ and set $\log w^{(i)}\gets0$
  \EndIf
  \State \textit{move} (optional): apply MCMC moves that leave $\gamma_{k+1}$ invariant
\EndFor
\State $\log\hat Z\gets\log\hat Z+\operatorname{logsumexp}_i\log w^{(i)}-\log n$
\State \Return samples $y_K^{(i)}$ with weights $w^{(i)}/\sum_jw^{(j)}$, and the free energy $-k_\mathrm BT\log\hat Z$
\end{algorithmic}
\end{algorithm}

\subsection{Inference-time steering}
\label{si:steering}
Steering samples a tilted version of the ensemble,
\begin{equation}
  p_{\lambda,r}(y)=\frac{\tilde p_\lambda(y)\,e^{r(y)}}{Z_{\lambda,r}},
  \label{eq:tilt}
\end{equation}
without retraining~\cite{cheng2026atlas,he2026rne}, where $r$ is a reward on the terminal configuration; for
Fig.~\ref{fig4}e it is a harmonic umbrella restraint on the height of one hydrogen atom (Section~\ref{si:pmf}). In
contrast to prior work~\cite{he2026rne,cheng2026atlas,blessing2026janus}, which tilts the learned distribution of the sampler, we tilt the physical
ensemble, so that the weights also correct the errors of the model and $Z_{\lambda,r}$ is an absolute normalizer. We run
Algorithm~\ref{alg:smc} with tilted intermediate targets, a guided proposal and weights that correct for both.

\paragraph{Intermediate reward.} An intermediate reward $r_t$ with $r_1=r$ tilts the intermediate targets into
$\gamma_k\,e^{r_{t_k}}$, and the first population, drawn from $\pi_0$, carries the weights $e^{r_0(y_0)}$. The values of
$r_t$ for $t<1$ do not change the target; they set how the tilt is introduced along the path. For independent
endpoints, the interpolant to the tilted ensemble has the marginals $\propto p_t(y)\,\mathbb E[e^{r(y_1)}\mid y_t=y]$, 
so we take $r_t(y)\approx\log\mathbb E[e^{r(y_1)}\mid y_t=y]$. For a reward on a collective variable $\xi$, we treat
$\xi(y_1)$ given $y_t$ as Gaussian with variance $v_t$ around $\hat\xi_t=\xi(\hat y_1)$, where
$\hat y_1=\mathrm{wrap}\big(y_t+[(1-\alpha)/\dot\alpha]\,b_\theta(y_t,t)\big)$ is the posterior-mean terminal implied by the
velocity~\cite{robbins1992empirical,cheng2026atlas}, so that
\begin{equation}
  r_t(y_t)=\log\int\mathcal N(\xi;\,\hat\xi_t,\,v_t)\,e^{r(\xi)}\,\mathrm d\xi .
  \label{eq:si-doob}
\end{equation}
The variance $v_t$ is estimated from untilted trajectories and vanishes at $t=1$, so that $r_1=r$; for the harmonic
restraint of Section~\ref{si:pmf}, Eq.~\eqref{eq:si-doob} is a Gaussian integral with a closed form.

\paragraph{Guided proposal and weights.} The proposal $\overrightarrow{q}_k$ adds the guidance $\varepsilon(t_k)\,u_k$,
with $u_k\approx\nabla r_{t_k}$, to the drift of the pretrained kernel $\overrightarrow{p}_{\theta,k}$ of
Eq.~\eqref{eq:em} and keeps its variance. Any $u_k$ is admissible, since the weights correct for it; for the umbrella
windows, $u_k$ differentiates $r_t$ through $\hat\xi_t$ only, without backpropagation through the network. After a step
$y_{k+1}\sim\overrightarrow{q}_k(\cdot\mid y_k)$, the log weight is incremented by
\begin{equation}
  \Delta\log w_k=\underbrace{\log\frac{\gamma_{k+1}(y_{k+1})}{\gamma_k(y_k)}
  +\log\frac{\overleftarrow{p}_{\theta,k}(y_k\mid y_{k+1})}{\overrightarrow{p}_{\theta,k}(y_{k+1}\mid y_k)}}_{\text{untilted}}
  \;+\;\underbrace{r_{t_{k+1}}(y_{k+1})-r_{t_k}(y_k)
  +\log\frac{\overrightarrow{p}_{\theta,k}(y_{k+1}\mid y_k)}{\overrightarrow{q}_k(y_{k+1}\mid y_k)}}_{\text{steering}}.
  \label{eq:smc-steer}
\end{equation}
The first part is the untilted increment of Eq.~\eqref{eq:smc}. The second is the steering increment, i.e., the change of the intermediate reward and the ratio of the pretrained to the guided
forward kernel. Since both kernels share the variance, this ratio is an explicit difference of squared residuals and vanishes without guidance.

With the initial weights $e^{r_0(y_0)}$, the increments telescope along a path to the path
weight of Eq.~\eqref{eq:logw} with the target $\tilde p_\lambda\,e^{r}$ and the proposal $\overrightarrow{q}_k$, from
which the learned energy and the intermediate reward cancel. Its expectation is $Z_{\lambda,r}$ for any intermediate
reward, guidance and step count, and with resampling the product of the mean weights remains an unbiased
estimate~\cite{delmoral2006sequential}.

\section{Oxide glasses}
\label{si:oxides}

\subsection{Reference data}
\paragraph{Interatomic potential.} The Bertani-Menziani-Pedone interatomic potential~\cite{bertani2021improved} assigns partial charges
(O $-1.2$, Li $0.6$, Ca $1.2$, Al $1.8$, Si $2.4$, P $3.0$\,e), which interact by damped shifted-force Coulomb summation
(damping parameter 0.25\,\AA$^{-1}$, cutoff 8\,\AA). Cation-oxygen and oxygen-oxygen pairs carry Morse terms with an
$r^{-12}$ repulsion and pairs of network formers a Buckingham repulsion, both truncated and shifted at 7\,\AA;
screened-harmonic three-body terms act on the angles between oxygen and the formers Si and P. 

\paragraph{Melt-quench reference.} For each composition, 32 independent runs start from uniformly random positions.
A random start places some atoms almost on top of each other. To remove these overlaps, the positions are first relaxed
by 300 steps of steepest descent in which no atom moves by more than 0.05\,\AA{} per step, followed by
10\,ps of Langevin dynamics at 4000\,K. The melt is then equilibrated for 100\,ps at 4000\,K,
quenched linearly at 5\,K\,ps$^{-1}$ to 1200\,K, held for 100\,ps and sampled for 200\,ps with one frame per
picosecond, using the BAOAB Langevin integrator with a time step of 1\,fs and a damping time of 0.1\,ps.

\subsection{Experimental setup}
\paragraph{Sampler setup.} The sampler generates the fractional positions on the torus at fixed species: the composition fixes the element of every atom index, in a random order per sample. The prior is uniform over the cell, and prior and terminal atoms of the same element are paired by permutation matching (Section~\ref{si:coupling}). The sampler draws the composition of every training configuration from the uniform
Dirichlet distribution over the mole fractions of the five oxides and rounds it to the nearest composition of whole
formula units with $N=255$ and at least one unit of each oxide. The composition enters the network as the normalized
atom fractions of the six elements, each with two sine-cosine features.

\paragraph{Network and generative process.} The network is a periodic PaiNN~\cite{schutt2021equivariant} with 4
layers, 64 channels, 16 Gaussian radial basis functions and a cutoff of 5\,\AA{}. Node
inputs are Fourier features of time, a 16-dimensional embedding of the element and the composition features. A gated
equivariant head emits the velocity $b_\theta$ and a scalar head the potential $\varphi_\theta$, whose gradient is the
score. Positions follow the interpolant of Eq.~\eqref{eq:si-interpolant} with $\alpha(t)=t$,
$x_t=\mathrm{wrap}(x_0+t\Delta)$, with the velocity target $\Delta$ and the score target of Section~\ref{si:score}, which for the uniform prior is the force at the terminal. Generation uses 64 Euler-Maruyama steps with a Langevin strength that decreases log-linearly from $10^{-1}$ to $10^{-4}$.

\paragraph{Training.} Training is data-free and uses the regression losses only without the consistency loss. Each
outer iteration generates 20 configurations with an exponential moving average (EMA) of the network weights (decay 0.995), labels them with forces, adds them to a first-in-first-out buffer with capacity 2000 and takes 50 AdamW steps on batches of 20 (gradient clipping at 1). Stage 1 runs 600
outer iterations of a cosine schedule (learning rate $3\times10^{-4}$);
stage 2 continues from it for 1000 outer iterations at $10^{-4}$.

\paragraph{Inference.} We draw 32 samples per composition with 64 generation steps and evaluate no energy at
inference. After each of the first 63 generation steps, the sampler takes one Hamiltonian Monte Carlo trajectory of 8
leapfrog steps that leaves $e^{\varphi_\theta(\cdot,t_k)}$ invariant at the time $t_k=k/64$ reached by step $k$, with
a step size adapted per configuration; the samples are not refined further
at $t=1$.

\section{Phase diagrams of SiO$_2$ and Ti}
\label{si:npt}

\subsection{Reference data}
The reference Gibbs free energy come from non-reversible parallel tempering~\cite{syed2022nonreversible} on the same
potentials and supercells as the samplers, with one ladder per phase and pressure (0, 4, 8 and 12\,GPa for SiO$_2$,
0-20\,GPa in steps of 4\,GPa for Ti; 30 ladders). A scan consists of two Metropolis-adjusted Langevin sweeps per
chain, over the positions and then over the six cell coordinates, with adapted step sizes, followed by deterministic even-odd swaps between neighboring chains. 
MBAR~\cite{shirts2008statistically} over all chains gives $G$ every 50\,K. 

\subsection{Experimental setup}
\paragraph{Supercells and priors.} Each phase is a supercell with $N=216$ atoms. Its unit cell is first relaxed with the potential, positions and
cell together, at a series of pressures across the window. The supercell is an integer multiple of this unit cell, so it keeps the
crystal structure of the phase, which is not cubic for most phases. Among all such multiples with 216 atoms, we take the
one whose box is closest to a cube, found with ASE~\cite{larsen2017atomic}, because it keeps the periodic images of an
atom farthest apart.
SiO$_2$ uses MACE-MH-1 and Ti uses MACE-MPA-0.
The prior of a phase is fitted to flexible-cell Metropolis-adjusted Langevin Monte Carlo with
the same potential on a grid of temperatures and pressures across the window and interpolated in $T$ and
$P$: an Einstein crystal centred on the mean sites with the measured per-species root-mean-square
displacements as widths, and a Gaussian over the six cell coordinates with the measured mean and spread.
Both are made slightly wider than measured, so that the prior covers the fluctuations of the target.

\paragraph{Network and generative process.} The network is a periodic PaiNN~\cite{schutt2021equivariant} with 4
layers, 96 channels, 20 Gaussian radial basis functions and a cutoff of 5\,\AA, evaluated on the distances in the
current cell. Node inputs are 16 Fourier features of time, a 16-dimensional species embedding, the cell coordinates
and the conditions $T$ and $P$, each normalized over the window with four sine-cosine pairs. Positions and cell coordinates follow the interpolant of Eq.~\eqref{eq:si-interpolant} with $\alpha(t)=1-(1-t)^3$,
and generation is Euler-Maruyama with 200 steps and a Langevin strength $\varepsilon=m\,k_\mathrm BT$, with
$m=1$\,\AA$^2$\,eV$^{-1}$ for the positions and $0.01$\,eV$^{-1}$ for the cell coordinates.

\paragraph{Training.} Stage~1 trains from scratch with the regression losses: each outer iteration generates 256
configurations at conditions drawn uniformly from the window with the EMA weights (decay 0.995), labels them with
energy, forces and stresses, adds them to a first-in-first-out buffer of 4096, and takes 100 AdamW steps on batches
of 32 (learning rate $3\times10^{-4}$, 300 warm-up steps, cosine decay, gradient clipping at 1); terminals with forces
above 30\,eV\,\AA$^{-1}$ are not used as labels. Stage~2 continues from the stage-1 weights and buffer at a learning
rate of $10^{-4}$ with 64 configurations per outer iteration at eight conditions, draws Ti conditions only up to
1600\,K, and adds the simulation-free consistency loss (Section~\ref{si:consistency}, $w=1$).

\subsection{Gibbs free energy and phase boundaries}
\paragraph{Free energy.} For every phase, we estimate the Gibbs free energy per atom $G(T,P)$ on the grid of the
reference, every 50\,K and 4\,GPa, with Algorithm~\ref{alg:smc}. We use 256 particles and 800 Euler-Maruyama steps, four times as many
as in training. The particles are resampled when the effective sample size falls below one half, and every
resampling is followed by two Metropolis-adjusted Langevin moves on the intermediate target.

\paragraph{Cell fluctuations.} In a finite supercell, $G$ contains a contribution from the thermal fluctuations of the
six cell coordinates. This contribution is a finite-size effect and is not the same for every phase, so we remove it
before comparing phases. With $\Sigma_c$ the covariance of the six cell coordinates over the generated samples, we use
\begin{equation}
  G^\ast=G+\frac{k_\mathrm BT}{2N}\log\det(2\pi\Sigma_c),
\end{equation}
which is the free energy of the phase with its cell held at the most probable volume and shape, if the cell
coordinates are Gaussian distributed. The correction lowers $G$ by 3-18\,meV/atom and changes the difference between
two phases by at most 3\,meV/atom. We apply it to the reference in the same way.

\paragraph{Phase boundaries.} Between grid points, we interpolate $G^\ast$ of each phase with a cubic Hermite
polynomial in $P$, whose slopes are given by $\partial G/\partial P=V$ with $V$ the mean volume per atom, and linearly
in $T$. At each condition the stable phase is the one with the lowest $G^\ast$, and a phase boundary is where the two
lowest values are equal. The boundaries of OMNIA and of the reference are obtained in the same way.


\section{Phase diagrams of Cu-Ag and Cu-Ni}
\label{si:alloys}
\subsection{Reference data}
Cu-Ag and Cu-Ni are described by the embedded-atom potentials of Williams et al.~\cite{williams2006embedded} and
Fischer et al.~\cite{fischer2019systematic}, truncated at 5.3 and 5.0\,\AA, with $N=108$ atoms at $P=0$. The reference
is semi-grand-canonical Monte Carlo with replica exchange in temperature, run separately for the solid and the liquid
on $7\times57$ states for Cu-Ni (1300-1900\,K, $\Delta\mu=0.10$-1.50\,eV) and $10\times41$ states for Cu-Ag
(600-1500\,K, $\Delta\mu=0.35$-0.95\,eV), with four walkers and 20\,000 sweeps per state. The semi-grand potential of
each phase follows along every isotherm from MBAR~\cite{shirts2008statistically}, up to a constant that is fixed by the
free energy of a pure element (from a harmonic reference crystal for the solids and thermodynamic integration from the
ideal gas for the liquids). OMNIA uses the same constant. Two further calculations give independent estimates: fixed compositions connected by single-site
substitutions and, for the Cu-Ag liquidus, a joint reweighting over all temperatures. The reference boundaries in
Fig.~\ref{fig3}a,e are the mean over these estimates, with their spread as uncertainty. The Cu-Ag solvus comes from the
fixed-composition calculation alone, since the semi-grand simulations do not cross the miscibility gap below 1000\,K.

\subsection{Experimental setup}
\paragraph{Experts and priors.} Table~\ref{tab:si-alloys} lists the experts. The solids start from an Einstein crystal
on the fcc sites, whose width is proportional to $\sqrt T$ for Cu-Ni and
equal to the root-mean-square displacement of the pure host element for Cu-Ag. The liquids start from uniformly
distributed positions. The prior of the logarithmic volume is a Gaussian around a volume interpolated between the pure
elements, fitted to short isobaric Monte Carlo runs with the same potential.

\begin{table}[h]
\centering
\small
\begin{tabular}{llcccccc}
\toprule
 & Expert & $T$ (K) & $\Delta\mu$ (eV) & Position prior & Steps $K$ & $\varepsilon$ & Iterations \\
\midrule
Cu-Ni & fcc & 1350-1750 & 0.25-1.25 & Einstein crystal & 108 & $10^{-3}\to10^{-5}$ & 400 \\
Cu-Ni & liquid & 1350-1750 & 0.25-1.25 & uniform & 432 & $1\to10^{-2}$ & 220 \\
Cu-Ag & Cu-rich fcc & 600-1300 & 0.35-0.74 & Einstein crystal & 108 & $10^{-3}\to10^{-5}$ & 400 \\
Cu-Ag & Ag-rich fcc & 600-1300 & 0.59-0.95 & Einstein crystal & 108 & $10^{-3}\to10^{-5}$ & 400 \\
Cu-Ag & liquid & 950-1450 & 0.35-0.95 & uniform & 1296 & $1\to10^{-2}$ & 300 \\
\bottomrule
\end{tabular}
\caption{Experts of the alloy phase diagrams with their training windows and settings. The Langevin strength
$\varepsilon$ decreases along the generative path and is given in fractional units; iterations are outer iterations.}
\label{tab:si-alloys}
\end{table}

\paragraph{Network.} The network is the PaiNN of Section~\ref{si:npt} with the cutoff of the potential, and takes the
logarithmic volume and $\Delta\mu$ in place of the cell coordinates and $P$. It returns the velocities of the positions
and the volume, a scalar $\varphi_\theta$ whose derivatives give the two scores, and a prediction of the energy changes
$\Delta U_{i\to k}$, from which the species posterior follows as in Eq.~\eqref{eq:species-loss}. This prediction is a
constant per element, measured on a randomly occupied fcc lattice, plus a per-site output that is initialized at zero,
so that the untrained sampler generates an ideal solution. For the two Cu-Ag solids, the constant is measured at the
volume of the host element (12.4 and 18.0\,\AA$^3$ per atom), so that the untrained samplers generate almost pure Cu
and almost pure Ag and start in their own phase.

\paragraph{Training.} Positions and volume use $\alpha(t)=1-(1-t)^3$ and the targets of Section~\ref{si:score}. The
endpoints are independent, except for the Cu-Ag liquid, which uses the permutation coupling of
Section~\ref{si:coupling} with $\epsilon_m=0.03$. Each outer iteration generates 384 configurations with the EMA weights
(decay 0.995), labels them with the energy, the forces, the volume derivative and the $N$ substitution energies, adds
them to a buffer of 8192 and takes 250 AdamW steps on batches of 48 (learning rate $3\times10^{-4}$, 500 warm-up steps,
cosine decay to 5\%, gradient clipping at 1). The loss weights of the velocity and the score are 250 and 80 for the
solids and 2.5 and 30 for the liquids. The volume is held at its prior draw for the first 60 outer iterations and is
then released, with its Langevin strength raised from $10^{-4}$ to $2\times10^{-3}$ over 80 iterations; the Cu-Ni liquid
keeps its prior draw throughout.

\paragraph{Inference.} Each expert generates 256 configurations per state with its training settings; the Cu-Ni fcc
expert uses 432 steps and a Langevin strength from $10^{-2}$ to $10^{-4}$. Cu-Ni is sampled on 18 isotherms from 1365
to 1765\,K over the $\Delta\mu$ grid of the reference, which extends beyond the training window to the states of pure
Ni. Cu-Ag is sampled every 15\,meV, the solids on 23 isotherms from 600 to 1300\,K and the liquid on 17 isotherms from
900 to 1300\,K.

\subsection{Phase boundaries}
\label{si:alloys-boundaries}
At fixed $T$, the states of a phase differ only in the factor $e^{\beta\Delta\mu N_\mathrm B}$, so that MBAR on the
generated compositions of an isotherm gives $\Phi_\alpha$ up to a constant and $\langle N_\mathrm B\rangle_\alpha$ on a
fine grid of $\Delta\mu$. Configurations that left the phase of their expert, identified by their root-mean-square
displacement from the lattice sites, are discarded. The constant follows from the states at which the expert generates
a pure element,
\begin{equation}
  \Phi_\alpha(T,\Delta\mu)=G^\mathrm A_\alpha(T)+k_\mathrm BT\log p_\alpha(N_\mathrm B=0\mid T,\Delta\mu),
\end{equation}
or its analogue at $N_\mathrm B=N$, with the pure-element free energy $G^\mathrm A_\alpha$ of the reference: pure Ni
for Cu-Ni, the host element for the Cu-Ag solids and pure Ag for the Cu-Ag liquid. Path weights are not used, since
their effective sample size is too small for the liquids. 

At each temperature, the stable phase has the lowest potential, and the boundaries are the crossings between stable
phases: the liquidus where the liquid crosses a solid, and the solvus where the two solids cross. The eutectic is the
temperature at which the liquid first becomes stable, interpolated between isotherms. Beyond the last state at which a
Cu-Ag fcc expert generates its own phase, its potential is continued linearly. Two parts of the Cu-Ag diagram are not
drawn: the liquidus above 1275\,K, where one of the liquid experts generates too few pure-Ag configurations, and the
solvus at 950-1000\,K, where the composition of the Cu-rich expert at the crossing is spread across the miscibility
gap.

\section{Hydrogen on Pt(111) and Pd(111)}
\label{si:surfaces}
\subsection{Reference data}
\paragraph{Slabs, potential and chemical potential.}
Pt(111) and Pd(111) are $4\times4\times6$ slabs built with ASE~\cite{larsen2017atomic} (lattice parameters 3.92 and
3.93\,\AA) with 96 metal atoms held fixed,
periodic in the surface plane, with 12\,\AA{} of vacuum. Hydrogen is confined to a domain $D$ of volume $V$ that spans
the lateral cell and the heights $[-2.5,5.0]$\,\AA{} (Pt) or $[-10.0,5.0]$\,\AA{} (Pd) relative to the top metal plane.
For Pt, $D$ contains the surface and the first subsurface layer; for Pd, it reaches through four interlayer gaps. The
energy of $N$ hydrogen atoms is their binding energy to the clean slab,
$U_N=E(\text{slab}+N\,\mathrm H)-E(\text{slab})$, from MACE-MH-1~\cite{batatia2025cross}. Configurations with an
H-metal distance below 0.5\,\AA{} or an H-H distance below 1.0\,\AA{} are excluded, $U_N=+\infty$. The H-H limit
excludes the H$_2$ molecule (bond length 0.75\,\AA), so the ensemble is that of atomic hydrogen. The chemical potential
is given relative to the H$_2$ molecule in the same model, $\mu_\mathrm H=\mu_0+\Delta\mu_\mathrm H$ with
$\mu_0=\tfrac12E_{\mathrm H_2}+E_\mathrm H=-3.3831$\,eV, where $E_{\mathrm H_2}=-4.5254$\,eV is the energy of the isolated H$_2$ molecule at its equilibrium bond length, relative to two isolated atoms, and
$E_\mathrm H=-1.1204$\,eV the energy of the isolated atom in the model. 

\paragraph{Pt(111).} The reference is grand-canonical Monte Carlo on this slab and domain with the same potential, but
with an H-H limit of 0.3\,\AA. Hydrogen atoms are inserted, deleted and displaced one at a time~\cite{adams1975grand}, and replica
exchange connects 16 chemical potentials $\Delta\mu_\mathrm H$ between $-0.65$ and $-0.10$\,eV. We run 10 independent
simulations of 8000 rounds (two at 300\,K, four each at 600 and 900\,K) and discard the first 2000 rounds. The
weighted histogram analysis method~\cite{kumar1992weighted} over the histograms of $N$ at all chemical potentials gives
the relative weight of every particle number, and from it $\Delta G_\mathrm{diff}$ (defined in
Section~\ref{si:slots}).

\paragraph{Pd(111).} The reference is a Gibbs sampler on the slots and ghost atoms that the sampler also uses
(Section~\ref{si:slots}), so that the number of hydrogen atoms fluctuates between 0 and $M=32$. One sweep redraws every
ghost atom from its density $g$, redraws the occupation of every slot from its conditional distribution at the given chemical potential,
and takes five Metropolis-adjusted Langevin steps of the hydrogen positions. Every state is run from an empty and from
a filled slab, with 64 to 128 walkers and 80 to 300 sweeps, and is accepted when the two runs agree.

\paragraph{Potential of mean force.} The reference is umbrella sampling with the same potential, slab, domain and
windows as in Section~\ref{si:pmf}. One round consists of a hybrid Monte Carlo move of all hydrogen
atoms~\cite{duane1987hybrid}, two moves that relocate an untagged atom to another surface site, and a replica exchange
between neighboring windows~\cite{sugita1999replica}. Each window runs 200 rounds of burn-in and 1500 rounds of
sampling, and the windows are combined with the weighted histogram analysis method.

\subsection{Experimental setup}
\paragraph{Free energy, slots and ghosts.}\label{si:slots}
With $Z_N$ of the Methods, we define the reduced free energy $a_N=\log Z_N-N\log V$, with $a_0=0$, and write the
weight of Eq.~\eqref{eq:gc-weight} as
\begin{equation}
  p_\lambda(N)\propto\exp\Big[a_N+N\log\frac{e^{\beta(\mu_0+\Delta\mu_\mathrm H)}V}{\Lambda^3}-\log N!\Big].
\end{equation}
The values $a_N$ at one temperature therefore give the isotherm at every $\Delta\mu_\mathrm H$. The differential
adsorption free energy is the chemical potential, relative to $\tfrac12\mathrm H_2$, at which $N$ and $N-1$ hydrogen
atoms are equally likely,
\begin{equation}
  \Delta G_\mathrm{diff}(N)=-k_\mathrm BT\Big[a_N-a_{N-1}+\log\frac{V}{N\Lambda^3}\Big]-\mu_0.
\end{equation}

To give every $N$ the same state space, the sampler acts on $M$ slots, with $M=24$ for Pt and $M=32$ for Pd. The first
$N$ carry hydrogen, and the remaining
$M-N$ carry ghosts, which do not interact and follow a fixed normalized density $g$ on $D$. Relative to the uniform
probability measure on $D^M$, the target for $N$ atoms is
\begin{equation}
  \tilde p_\lambda(\mathbf x\mid N)=e^{-\beta U_N(\mathbf x_{1:N})}\prod_{i=N+1}^{M}V\,g(\mathbf x_i),\qquad
  \int\tilde p_\lambda(\mathbf x\mid N)\,\frac{\mathrm d^{3M}x}{V^M}=e^{a_N}.
  \label{eq:si-slot-target}
\end{equation}
The ghosts integrate to one, so they change neither $Z_N$ nor the distribution of the hydrogen atoms.

\paragraph{Site prior.}\label{si:site-prior}
The density $g$ is also the prior of every slot. It places hydrogen near the binding sites of a single atom and is
built from the potential alone. We evaluate the energy of one hydrogen atom on a grid over one $1\times1$ surface cell,
select the lowest points that lie within 0.25\,eV (Pt) or 0.55\,eV (Pd) of the minimum and at least 0.8\,\AA{}
apart, at most four for Pt and eight for Pd, and replicate them over the 16 surface cells. The density $g$ is a mixture
of Gaussians on these points (widths 0.4\,\AA{} in the surface plane and 0.3\,\AA{} along the surface normal), with
total weight 0.7 (Pt) or 0.98 (Pd), and a uniform density on $D$ with the remaining weight.

\paragraph{Network and training.}
The network is a PaiNN-type equivariant message-passing network~\cite{schutt2021equivariant} on the $M$ slots and the
metal atoms of the top three (Pt) or six (Pd) layers, with 4 layers, 96 channels, 20 radial basis functions and a cutoff
of 5.5\,\AA. Each slot receives the time, a token that marks it as hydrogen or ghost, the conditions $T$ and $N$, and
its height above the surface; ghosts send no messages. The network returns the velocity and the score of every slot.
The endpoints are independent for Pt and use the permutation coupling of Section~\ref{si:coupling} for Pd. The
generative process uses $K=400$ steps with a Langevin strength of $\varepsilon=0.2$\,\AA$^2$ for Pt, and $K=200$ steps
with $\varepsilon=0.2\,\text{\AA}^2\times T/300\,\mathrm K$ for Pd.

Each outer iteration draws $T$ uniformly from the window (300-900\,K for Pt, 300-600\,K for Pd) and $N$ uniformly up
to the largest $N$ at that temperature. The largest $N$ is 20, 16 and 12 at 300, 600 and 900\,K for Pt, and 28 and 32
at 300 and 600\,K for Pd, interpolated linearly in between. The iteration then generates configurations with the EMA
weights (decay 0.995), labels them with energy and forces (clipped at 20\,eV\,\AA$^{-1}$ per atom), adds them to a
first-in-first-out buffer of 2048, and takes 100 Adam steps (learning rate $3\times10^{-4}$, 200 warm-up steps, cosine
decay to 20\%, gradient clipping at 1). We train three models per metal with different seeds for 3000 outer iterations,
with the regression loss only.

\paragraph{Evaluation.}
For every temperature and every $N$, the sampler draws $n$ paths with $K=800$ steps, and the log weights of
Eq.~\eqref{eq:logw} with the target of Eq.~\eqref{eq:si-slot-target} give
$\hat a_N=\operatorname{logsumexp}_i\log w_i-\log n$ at one energy evaluation per path. The estimates of $Z_N$ are
averaged over the three models, for Pd over two checkpoints of each model. From $\hat a_N$ follow the
probabilities $p_\lambda(N)$ of the particle numbers and the coverage
$\langle N\rangle=\sum_NN\,p_\lambda(N)$ at every $\Delta\mu_\mathrm H$, without further sampling.

\subsection{Potential of mean force by steered sequential Monte Carlo}
\label{si:pmf}
Fig.~\ref{fig4}e is obtained by inference-time steering (Section~\ref{si:steering}) at 600\,K and $N=16$ (1\,ML).
Steering uses the learned density $e^{\varphi_\theta}$ as intermediate target, so we use a Pd model that is also trained
with the consistency loss of Section~\ref{si:consistency}: 600 outer iterations with the regression loss, followed by
1000 with both losses. 

\paragraph{Windows.} The collective variable is the height $\xi(\mathbf x)$ of one tagged hydrogen atom above the top Pd
plane. Since the hydrogen atoms are exchangeable, tagging one atom needs no change to the model. Window $k$ is the
tilted ensemble of Eq.~\eqref{eq:tilt},
\begin{equation}
  p_k(\mathbf x)=\frac{\tilde p_\lambda(\mathbf x\mid N)\,e^{r_k(\mathbf x)}}{Z_k},\qquad
  r_k(\mathbf x)=-\tfrac12\beta\kappa\,\big(\xi(\mathbf x)-\xi_{0,k}\big)^2,
\end{equation}
with $\kappa=3$\,eV\,\AA$^{-2}$ and centres $\xi_{0,k}$ from $-5.0$ to 2.0\,\AA{} in steps of 0.2\,\AA. Each window is
sampled by steered sequential Monte Carlo, following ATLAS~\cite{cheng2026atlas} (Section~\ref{si:steering}), with
$n=2048$ particles, $K=800$ steps and $\varepsilon=0.8$\,\AA$^2$. The variance $v_t$ of the intermediate reward is
measured on unsteered trajectories and multiplied by three, since the measured value makes the reward too narrow early
in the path and the particle population collapses. The potential is evaluated only over the last 5\% of the path, where
the learned target is blended into the true one.

\paragraph{Potential of mean force profile.} Steered sequential Monte Carlo returns weighted samples of $p_k$ together with an
estimate of $Z_k$, so that every window gives the density of the height on an absolute scale,
\begin{equation}
  \rho(\xi)=Z_k\,\mathbb E_{p_k}\big[\delta\big(\xi(\mathbf x)-\xi\big)\,e^{-r_k(\mathbf x)}\big],
  \qquad F(\xi)=-k_\mathrm BT\log\rho(\xi).
\end{equation}
The windows therefore need no relative constants. We estimate $\rho$ from the weighted samples of each window in bins of
0.05\,\AA{} and average $\log\rho$ over the windows, weighted by their effective sample counts. For the comparison in
Fig.~\ref{fig4}e, the profile is shifted by one constant onto the reference.

\section{Sodium in amorphous carbon}
\label{si:sodium}

\subsection{Reference data}
\paragraph{Frameworks and potential.} Carbon and sodium are described by the GAP of Front
et al.~\cite{front2026density}. A framework is
generated from $N_\mathrm C$ carbon atoms at random positions in a cubic cell of edge $L=17.08$\,\AA, which holds 500
atoms at 2.0\,g\,cm$^{-3}$, by a melt-quench with Langevin dynamics: 2\,ps at 5000\,K, 20\,ps at 3500\,K, a quench to
300\,K at 1000\,K\,ps$^{-1}$ and a final relaxation. We run 32 melt-quenches for every $N_\mathrm C$ from 250 to 500.
The model is evaluated on 50 frameworks, 8 or 9 at each of $N_\mathrm C=250,300,\dots,500$. Two of them were never
used in training; the others were available to it.

\paragraph{Grand-canonical Monte Carlo.} For each of the 50 frameworks, the reference is GCMC~\cite{frenkel2023understanding} with replica exchange
between 16 chemical potentials from $-1.42$ to $-0.97$\,eV, in 16 independent copies started from the empty framework.
One iteration of the simulation consists of a Metropolis-adjusted Langevin move of all sodium atoms, four insertion or deletion attempts, one
relocation of a random atom, and an exchange between neighboring chemical potentials~\cite{sugita1999replica}.
Insertions are proposed near the binding sites of a single sodium atom, next to sodium atoms that are already present,
or uniformly in the cell. Each chain runs 6000 iterations, and the bound sodium is read from
the final states of the 16 chains at $\mu_\mathrm{Na}=-1.24$\,eV. 

\subsection{Experimental setup}
\paragraph{Prior.} The prior places sodium on the binding sites of the framework without evaluating the GAP. It is
based on an effective energy of a single sodium atom,
\begin{equation}
  \epsilon(\mathbf r)=\epsilon_\infty+\sum_j\phi(|\mathbf r-\mathbf r_j|)+\epsilon_\mathrm{pore}\,s\big(d_\mathrm C(\mathbf r)\big),
\end{equation}
where the sum runs over the carbon atoms. The pair potential $\phi$, which vanishes beyond 6\,\AA, and the offset
$\epsilon_\infty$ are fitted once to 640\,000 GAP energies of a single sodium atom at random positions in 16 frameworks.
The last term lowers the energy by $\epsilon_\mathrm{pore}=-0.7$\,eV in the interior of the pores, where $s$ rises
from 0 to 1 as the distance $d_\mathrm C$ to the nearest carbon atom grows from 3.5 to 4.5\,\AA. It stands in for the
cohesion of sodium clusters, which the energy of a single atom lacks. The binding sites are the points of lowest
$\epsilon$ on a $64^3$ grid, selected greedily at least 2.8\,\AA{} apart, which gives 116-161 sites per framework. The
prior is a lattice gas on these sites: every site holds at most one sodium atom, close to its centre, with a weight
given by the Boltzmann factor $e^{-\beta\epsilon}$ at 300\,K, and atoms without a site are placed in the remaining
volume. Its normalizer for $N$ atoms is computed exactly by dynamic programming over the sites, and its samples are
exact.

\paragraph{Network.} The network is a PaiNN-type equivariant message-passing network~\cite{schutt2021equivariant} in
two parts. A carbon encoder (3 layers, 96 channels, cutoff 4.0\,\AA) depends neither on time nor on the sodium atoms and
is evaluated once per framework at inference. Four sodium layers (96 channels, 20 radial basis functions, cutoff
5.0\,\AA) pass messages from neighbouring sodium atoms and from the encoded carbon atoms. The sodium atoms receive the
time and the conditions $N$ and $N_\mathrm C$; the temperature is fixed at 300\,K. The network returns the velocity of
every sodium atom and a scalar $\varphi_\theta$, whose gradient with respect to the sodium positions is the score.

\paragraph{Training.} The interpolant is that of Eq.~\eqref{eq:si-interpolant} with $\alpha(t)=t$ and the permutation
coupling of Section~\ref{si:coupling}, and the generative process uses $K=200$ steps with $\varepsilon=0.05$. Training
uses the regression loss only. Each of the 200 outer iterations generates 192 configurations, for each of which
$N_\mathrm C$, the framework and $N\le200$ are drawn uniformly. They are labelled with GAP energy and forces (clipped
at 20\,eV\,\AA$^{-1}$ per atom) and added to a first-in-first-out buffer of 2048, followed by 100 AdamW steps on
batches of 48 (learning rate $3\times10^{-4}$, 200 warm-up steps, gradient clipping at 1, EMA decay 0.995).

\subsection{Free energy and bound sodium}
\paragraph{Free energy.} As in Section~\ref{si:slots}, $a_N=\log Z_N-N\log V$, with $V=L^3$. For each framework, $a_N$ is estimated
at $N=1,10,20,\dots,80$ by sequential Monte Carlo (Algorithm~\ref{alg:smc}) with $n=256$ particles along the $K=200$
steps of the learned path, with the learned $\varphi_\theta$ as intermediate targets and one Metropolis-adjusted
Langevin move per step, which requires no GAP evaluation. Then 50 further steps follow, in which the target is changed
linearly from the learned $\varphi_\theta(\cdot,1)$ to $-\beta U_N$, each with a reweighting, resampling if needed and
one Metropolis-adjusted Langevin move with GAP forces. The values at all other $N$ follow from a cubic polynomial fitted to
the increments $a_N-a_{N-1}$, and the weight of Eq.~\eqref{eq:gc-weight} is
\begin{equation}
  p_\lambda(N)\propto\exp\Big[a_N+N\log\frac{e^{\beta\mu_\mathrm{Na}}V}{\Lambda^3}-\log N!\Big],\qquad N\le80.
\end{equation}

\paragraph{Bound sodium.} At low carbon density, $\mu_\mathrm{Na}=-1.24$\,eV lies close to the filling of the large
pores by sodium clusters, whose atoms are held by other sodium atoms rather than by the carbon. Their number converges
slowly: at 1.0 and 1.2\,g\,cm$^{-3}$, GCMC runs started from empty and from overfilled frameworks end about six sodium
atoms apart. We therefore count only the bound sodium atoms, whose nearest carbon atom is closer than 3.0\,\AA, the
first minimum of the distribution of nearest-carbon distances. From 1.4\,g\,cm$^{-3}$ upwards this removes less than one
atom per cell, and at 1.0\,g\,cm$^{-3}$ about a quarter of the atoms. The bound sodium and the pair distribution
functions are computed from two further runs, with the settings above, at the most probable count $N^\ast$ of each
framework. The distribution of the number of bound atoms in one framework then follows from
Eq.~\eqref{eq:bound-sodium}.

\bibliography{refs.bib}
\end{document}